\documentclass[titlepage,12pt]{article}
\usepackage{amssymb,amsmath,amsfonts}
\usepackage{graphicx}
\usepackage{cite}
\makeatletter
\def\@sect#1#2#3#4#5#6[#7]#8{\ifnum #2>\c@secnumdepth
  \def\@svsec{}\else
  \refstepcounter{#1}\edef\@svsec{\csname the#1\endcsname.\hskip0.5em}\fi
  \@tempskipa #5\relax
  \ifdim \@tempskipa>\z@
    \begingroup
      #6\relax
      \@hangfrom{\hskip #3\relax\@svsec}{\interlinepenalty \@M #8\par}%
    \endgroup
    \csname #1mark\endcsname{#7}\addcontentsline
      {toc}{#1}{\ifnum #2>\c@secnumdepth \else
        \protect\numberline{\csname the#1\endcsname}\fi #7}%
  \else
    \def\@svsechd{#6\hskip #3\@svsec #8\csname #1mark\endcsname
      {#7}\addcontentsline{toc}{#1}{\ifnum #2>\c@secnumdepth \else
        \protect\numberline{\csname the#1\endcsname}\fi #7}}%
  \fi \@xsect{#5}}

\newcommand{\tbart}{t{\bar t}}
\newcommand{\ttbar}{t{\bar t}}
\newcommand{\mtt}{M_{t\bar t}}

\renewcommand{\thefootnote}{\small\fnsymbol{footnote}}

\begin{document}
\begin{titlepage}
  \begin{flushright}
TTK-12-19 
    \end{flushright}
\vspace{0.01cm}
\begin{center}
{\LARGE {\bf Top quark and leptonic charge
    asymmetries for the Tevatron and LHC}  \\
\vspace{1.5cm}
\large{\bf Werner Bernreuther\,$^{a}$\footnote{\tt breuther@physik.rwth-aachen.de} and 
 Zong-Guo Si\,$^{b}$}\footnote{\tt zgsi@sdu.edu.cn}
\par\vspace{1cm}
$^a$Institut f\"ur Theoretische Physik, RWTH Aachen University, 52056 Aachen, Germany\\
$^b$Department of Physics, Shandong University, Jinan, Shandong
250100, China
\par\vspace{1cm}
{\bf Abstract}\\
\parbox[t]{\textwidth}
{\small{ We compute, for $\ttbar$ production at the LHC and at the Tevatron,
   several charge asymmetries to next-to-leading oder (NLO) QCD, including also the electromagnetic
   and weak-interaction corrections. We calculate these asymmetries 
  both inclusively and with
   additional kinematic cuts and compare our results, where possible,
   with recent experimental results and with Standard Model (SM) predictions.
    The $\ttbar$ asymmetries induce also corresponding asymmetries
    for the charged leptons from semileptonic top-quark decay. Although 
   these asymmetries are, in the SM,   smaller than the corresponding
   ones for top quarks,
    they are expected to be measurable quite  precisely. In fact,
  measurement of a lepton asymmetry in $\ell$ + jets events was
  reported by the D$\emptyset$  \cite{Abazov:2011rq} and CDF
  \cite{CDF2012} experiments.  We analyze and compute to NLO
   in the gauge couplings  leptonic charge asymmetries for dileptonic and
   semileptonic $\ttbar$ events, with and without acceptance cuts, at
   the Tevatron and the LHC. 

}}
}
\end{center}
\vspace*{1.5cm}

PACS number(s): 12.38.Bx, 13.88.+e, 14.65.Ha
\end{titlepage}
%
%
\setcounter{footnote}{0}
\renewcommand{\thefootnote}{\arabic{footnote}}
\setcounter{page}{1}

\section{Introduction} 
\label{introduction}

So far, almost all of the experimental results from the Tevatron and
the LHC on top quark production and decay imply that this quark
behaves pretty much as expected from the Standard Model -- the
exception being the measurements of the charge asymmetry in $\ttbar$
production  at the Tevatron by the CDF and D$\emptyset$ experiments
 reported in \cite{Aaltonen:2011kc,Abazov:2011rq}, which are (considerably)
 higher than   the
SM predictions \cite{Kuhn:1998jr,Kuhn:1998kw,Bowen:2005ap,Antunano:2007da}.
In particular,  a  $3.4 \sigma$ deviation from the SM was cited
  in \cite{Aaltonen:2011kc} for the CDF determination  \cite{Aaltonen:2011kc} of 
  the 
  $\ttbar$ rest frame  asymmetry
 for high pair-invariant mass, $A^{t\bar t}(M_{t\bar t} >  450~{\rm GeV})$.
  This triggered a very large
number of investigations on possible new physics contributions to
$\ttbar$ production. (For recent reviews, see for instance
\cite{Kamenik:2011wt,Westhoff:2011tq,AguilarSaavedra:2012ma}.)
A recent CDF measurement \cite{CDF2012} 
 of this observable based on a larger data  
  sample and, on the theory side,  the incorporation of
  the complete ${\cal O}(\alpha_s^2\alpha)$
  electroweak corrections  \cite{Hollik:2011ps} alleviated this tension as far
  as $A^{t\bar t}(M_{t\bar t} >  450~{\rm GeV})$ is concerned, 
  but did not remove it.

 The situation is  unclear for several reasons.
 The D$\emptyset$ experiment
at the Tevatron did not confirm \cite{Abazov:2011rq}  the
significant enhancement of 
    the $\ttbar$ rest-frame   asymmetry at high $\mtt$ seen by CDF.
 On the other hand,  D$\emptyset$ measured a leptonic
   charge asymmetry $A^\ell$ in $\ttbar \to \ell +$ jets \cite{Abazov:2011rq}, which is 
  considerably larger than the corresponding SM prediction 
 \cite{Bernreuther:2010ny}, while a recent  CDF 
 measurement \cite{CDF2012}, which is however not yet corrected for
  detector effects and acceptance,  agrees with it.
 Moreover, the $\ttbar$ charge asymmetries $A_C$ measured 
  by the CMS \cite{Chatrchyan:2011hk,CMSchargeas} and ATLAS \cite{Aad:2012ug}
 collaborations   at the LHC agree, within the present 
  uncertainties, with the
 SM calculations \cite{Kuhn:2011ri}. Obviously, it is of prime importance to explore and
 hopefully clarify this
 topic in detail in the (near) future,
 both by experiment and  theory. 

 As the SM-induced charge asymmetry  in $\ttbar$ production at the
 LHC is small,  a number of observables  related to $A_C$
   have been proposed and analyzed, including those in 
\cite{Antunano:2007da,Wang:2010du,Xiao:2011kp,Hewett:2011wz,Arguin:2011xm,AguilarSaavedra:2011cp,Kuhn:2011ri,Alvarez:2012vq,Alvarez:2012uh},
  that enhance the (predominantly QCD induced)
   effect and serve to discriminate between the SM and
   various  new physics models\footnote{An interesting proposal of
     ``collider independent'' charge asymmetries was recently made in
     \cite{AguilarSaavedra:2012va}.}. 
      In this paper we compute some of these
   asymmetries at next-to-leading order (NLO) in the SM gauge
   couplings. Here, this notation refers to the computation of  the numerators of the asymmetries 
   to  order $\alpha_s^3$ in the QCD coupling
   including the mixed QCD-QED and mixed QCD-weak interaction
   corrections. For some of these observables, respective results were
   recently obtained in the literature
   \cite{Hollik:2011ps,Wang:2010du,Xiao:2011kp,Kuhn:2011ri} with which we compare; some of
   our results are new. So far, most of the predictions were made
   at the $\ttbar$ production level, while  the experimental measurements
    of the charge asymmetries at the
   ``reconstruction level'' were unfolded,
   i.e., corrected for detector acceptance and resolution, to obtain
   the corresponding $\ttbar$ ``production level'' asymmetries.

   The $\ttbar$ asymmetries induce also corresponding asymmetries
    for the charged leptons from semileptonic top-quark decay. Although 
   these asymmetries are, in the SM,   smaller than the corresponding
   ones for top quarks \cite{Bernreuther:2010ny},
    they should be measurable more precisely. In fact,
  measurement of a lepton asymmetry in $\ell$ + jets events was
  reported by D$\emptyset$  \cite{Abazov:2011rq} and CDF \cite{CDF2012}.
  We analyze and compute to NLO
   in the gauge couplings  leptonic charge asymmetries for dileptonic and
   semileptonic $\ttbar$ events, with and without acceptance cuts, at
   the Tevatron and the LHC. This extends our previous results for the
   Tevatron \cite{Bernreuther:2010ny}.

 It seems appropriate to  briefly recapitulate here the status of the SM predictions concerning
 the $\ttbar$ charge asymmetries in hadronic production. In the SM,
 the leading-order effect is induced by the NLO QCD, i.e.,  the ${\cal
   O}(\alpha_s^3)$  contributions $d\sigma_{A,\ttbar}$ to the differential $\ttbar$ cross
 section  which are odd with respect
 to the exchange of $t \leftrightarrow \bar t$.   The first  dedicated
 NLO QCD prediction of the charge asymmetry, 
including an estimate of the electroweak contributions, was made in
\cite{Kuhn:1998jr,Kuhn:1998kw}.  Subsequent analyses were done in
\cite{Bowen:2005ap,Antunano:2007da}. 
 In \cite{Bernreuther:2010ny}   the mixed QCD-weak
   corrections of ${\cal O}(\alpha_s^2\alpha)$ were
   included. Ref. \cite{Hollik:2011ps} determined, besides the weak contributions,  also the 
QCD-QED  contributions  of ${\cal O}(\alpha_s^2\alpha)$  which are, in
fact, more important than  the weak-interaction corrections,  and
obtained predictions of the Tevatron asymmetries to NLO in the SM
gauge couplings; cf. also \cite{Kuhn:2011ri}.
 The NNLO QCD corrections to $d\sigma_{A,\ttbar}$ are not yet known\footnote{
 For $\ttbar$+jet events, QCD induces a charge asymmetry already at
 tree-level, which receives large corrections
 at NLO
 \cite{Dittmaier:2007wz,Dittmaier:2008uj,Melnikov:2010iu,Melnikov:2011qx,Alioli:2011as}. 
  Ref. \cite{Melnikov:2010iu}
 argues that the inclusive $\ttbar$ asymmetries may not receive such
  large QCD corrections beyond NLO QCD.}. The fixed-order NLO QCD computations were
   supplemented by soft-gluon resummation at next-to-leading (NLL) \cite{Almeida:2008ug} and
  next-to-next-to-leading (NNLL) \cite{Ahrens:2010zv,Kidonakis:2011zn,Ahrens:2011uf} logarithmic order. 
   These corrections do not alter  the   fixed-order NLO QCD results 
    significantly.
 The QCD-induced charge asymmetries can and are being computed also
 with the widely used NLO QCD Monte Carlo
 programs  \cite{Frixione:2008ym,Frixione:2007nw,MCFM,Campbell:2010ff,Campbell:2012uf}.
   An issue, which in the past has been a source of confusion between
  theorists and experimentalists, is how the asymmetries are 
  computed in the context of NLO Monte Carlo simulations, see Sect.~\ref{sec.rett}.
 
 The paper is organized as follows. In Sect.~\ref{sec.rett} we compute
 a number of $\ttbar$ charge asymmetries at NLO in the SM gauge
 couplings for the Tevatron and the LHC at 7, 8, and 14 TeV
 center-of-mass energy and compare, where possible,  with experimental
 results and other SM calculations.  In Sect.~\ref{sec:diljet}
  we make corresponding SM predictions for two leptonic asymmetries
  for dileptonic and lepton plus jets $\ttbar$ events at the Tevatron
  and for some  leptonic asymmetries for dileptonic final states at
  the LHC.   Sect.~\ref{sec:concl} contains a summary and outlook.

\section{Top-quark charge and   forward-backward asymmetries}
\label{sec.rett}

In this section   we consider various forward-backward and charge asymmetries
 for the Tevatron and LHC at the level of $\ttbar$ on-shell intermediate states
  and calculate these asymmetries within the SM.

\subsection{Tevatron}
 \label{suse:tevatron}
First  we compute the top-quark laboratory- and rest-frame
   charge/forward-backward
   asymmetries  for the Tevatron to NLO QCD including the photonic and weak-interaction
  contributions. The sole purpose of this section is to compare the results
  of our computational setup with previous SM computations of these
   asymmetries \cite{Kuhn:1998jr,Kuhn:1998kw,Bowen:2005ap,Antunano:2007da,Almeida:2008ug,Hollik:2011ps,Kuhn:2011ri,Bernreuther:2010ny,Ahrens:2010zv,Kidonakis:2011zn,Ahrens:2011uf,Campbell:2012uf}, and also 
 with recent experimental results \cite{Aaltonen:2011kc,Abazov:2011rq,CDF2012}.

For top-quark pair production at the Tevatron, $p {\bar p} \to \ttbar +X$,
 the differential and  integrated charge 
  asymmetry, 
$A(y)$ and $A$, are  defined by
\begin{equation}
 A(y) =  \frac{N(y_t) - N(y_{\bar t})}{N(y_t) + 
  N(y_{\bar t})} \, , \qquad  
   A =\frac{N(y_t>0) - N(y_{\bar t}>0)}{N(y_t>0) + N(y_{\bar t}>0)}\, ,
\label{chasy}
\end{equation}
where  $y_t$, $y_{\bar t}$ denote the rapidities of 
the $t$ and $\bar t$ quark  in the laboratory frame, 
  and $N(y)=d\sigma_{\tbart} /d y$. CP invariance implies that for $\ttbar$
 production at the Tevatron  $N(y_{\bar t}) = N(-y_t)$, which in turn  
 implies that  $A$
is equal to the forward-backward asymmetry of the top quark:
\begin{equation}
A_{FB}^t = \frac{N(y_t>0) - N(y_t<0)}{N(y_t>0) +
  N(y_t<0)}\, \quad \text{and} \quad A_{FB}^{\bar t}= - A_{FB}^{t}\, .
\label{fobasy}
\end{equation}
 Another important observable is the pair asymmetry or $\ttbar$ rest-frame asymmetry
\begin{equation} 
 A^{\tbart} = \frac{N(\Delta y > 0) - N(\Delta y <
       0)}{ N (\Delta y > 0 ) +  N(\Delta y  < 0)}\, ,
\label{fobattasy}
\end{equation}
where $\Delta y = y_{t} - y_{\bar t}$. As $\Delta y$ is boost-invariant
  along the beam axis, this rapidity difference is, in the limit of
  small $p_T$ of the $\ttbar$ system,  the
   same  in the hadronic and the $\ttbar$ rest frame. 
The sign of $\Delta y$ is, in fact, invariant under such a boost. 
The asymmetry (\ref{fobattasy}) is, 
  for kinematical reasons, larger than \eqref{fobasy}. \\

The Bose symmetry of the $gg$ state precludes a  contribution
to the  asymmetries $A$, $A_{FB}^t$, and $A^{\tbart}$ from
   $gg \to \ttbar X$ --
 irrespective of whether or not the production density matrix $R_{gg}$
  contains P- and/or CP-violating pieces.
  The asymmetries are generated by the
   interference of  even and odd terms 
   under $t\leftrightarrow
{\bar t}$  -- while the initial partons are kept fixed -- in the amplitudes
for $q {\bar q} \to \ttbar X$ and, likewise, for
$g q  \to \ttbar q \ (+X)$ and $g {\bar q} \to  \ttbar {\bar q} \ (+X)$. 

In the SM the dominant contributions to \eqref{chasy} - \eqref{fobattasy}
 arise from the NLO QCD corrections to $\ttbar$ production by
  $q {\bar q}$ annihilation, i.e., terms of order $\alpha_s^3$ in the 
  partonic cross section $d{\hat\sigma}(q {\bar q} \to \ttbar X)$
  which are antisymmetric under the interchange of $t$ and $\bar t$. 
These terms comprise, for  $q {\bar q} \to \ttbar$, the  antisymmetric part of
 the interference of the 
 Born diagram with the  1-loop
 box and crossed box diagrams and, for  $q {\bar q} \to \ttbar g$, the
 antisymmetric part of the interference of initial and final state
 radiation. In addition, antisymmetric interference terms   
  of order $\alpha_s^3$ in the squared matrix elements of $g q  \to
  \ttbar q$ and $g {\bar q} \to  \ttbar {\bar q}$, respectively,
  contribute  also to  the above asymmetries. At the Tevatron,
  they are numerically irrelevant, while at the LHC they may reach
  a sizeable fraction of the  contributions from  $q \bar q$
  annihilation (see below).

 As was pointed out in \cite{Hollik:2011ps}, the mixed QCD-QED contributions of
 order $\alpha_s^2 \alpha$ to the asymmetries  from
 $q {\bar q} \to \ttbar,\ \ttbar g, \ \ttbar \gamma$ are
 important.  (These corrections had been estimated previously
   in \cite{Kuhn:1998kw}.) At the level of the $q {\bar q}$ initial states, the
  ratio of the mixed QCD-QED and pure QCD
 contributions is $R_q=(36/5)Q_q Q_t\alpha/\alpha_s$ \cite{Hollik:2011ps}
 (where $Q_a$ denotes the charge of quark $a$ in units of $e$).
  For  $p {\bar p}$ collisions at the Tevatron this implies that
  the ratio of the  corresponding contributions to    \eqref{chasy} -
  \eqref{fobattasy} is about $18\%$. At the LHC this ratio decreases
  to about $13\%$ because the ratio of $u{\bar u}$ and $d{\bar d}$
  collisions decreases from about $4:1$ at the Tevatron to $2:1$ at
  the LHC.  
 
 The (nominally) leading effects of the weak interactions on the asymmetries are as
 follows. At Born level there is the contribution of ${\cal
   O}(\alpha^2)$ 
  from the antisymmetric terms of the squared
 amplitudes of $q{\bar q}\to \gamma, Z \to \ttbar$. Then there are
 antisymmetric terms in the mixed QCD-weak corrections of 
${\cal O}(\alpha_s^2\alpha)$ to  $q{\bar q}\to \ttbar (g)$. These are
contained in i) the interferences of the ${\cal O}(\alpha_s^2)$ two-gluon
box diagrams with the Born $Z$-exchange diagram and of  the ${\cal
  O}(\alpha_s\alpha)$ Z-gluon box diagrams with the Born gluon exchange
diagram, and ii) in the interferences of the  ${\cal  O}(g_s^3)$ and 
 ${\cal  O}(g_s e^2)$ gluon bremsstrahlung diagrams. (The contribution
 from $Z$ boson radiation, $q{\bar q}\to \ttbar Z$, to  the
 inclusive asymmetries is very small  and will be neglected.)  At the Tevatron
 and the LHC  these weak interaction corrections increase the QCD 
 asymmetries by a few  percent (cf. \cite{Bernreuther:2010ny,Hollik:2011ps,Kuhn:2011ri} and below). The weak
 interactions induce also parity-violating form factors at 1-loop in the $q{\bar
   q}g$ and $\ttbar g$ vertices; however, they make,  at ${\cal
   O}(\alpha_s^2\alpha)$, no contribution to
 the antisymmetric part of  $d{\hat\sigma}(q {\bar q} \to \ttbar X)$.
  There are also mixed QCD-weak contributions of ${\cal
   O}(\alpha_s^2\alpha)$ and  ${\cal O}(\alpha_s\alpha^2)$ to the
  asymmetries from  $g q \ ({\bar q}) \to  \ttbar q \ ({\bar q})$.
  They are negligibly small for the Tevatron, but  at the LHC they 
   are of comparable
  size  as the mixed QCD-weak contributions  to $q \bar q$
  annihilation (see Sect.~\ref{suse:lhc}).

 In the following we compute the asymmetries \eqref{fobasy} and
 \eqref{fobattasy}, taking into account in the numerators 
  the ${\cal O}(\alpha_s^3)$ QCD
 and the ${\cal O}(\alpha^2)$ and  ${\cal
   O}(\alpha_s^2\alpha)$ electroweak corrections as discussed above. 
 (As to the weak interaction corrections, we use our previous results
  \cite{Bernreuther:2005is,Bernreuther:2006vg,Bernreuther:2008md},
cf. also \cite{Kuhn:2005it,Kuhn:2006vh,Beenakker:1993yr}.)
 To this order,  a consistent fixed-order perturbative expansion of the
ratios  \eqref{fobasy}, \eqref{fobattasy} 
 precludes taking into account the  NLO QCD corrections to the
 denominators. Therefore we evaluate the denominators of all the
 asymmetries considered in this paper with  LO QCD
 matrix elements (as was done in \cite{Kuhn:1998jr,Kuhn:1998kw,Antunano:2007da,Hollik:2011ps,Kuhn:2011ri,Bernreuther:2010ny}). 
 As to the use of  parton distribution functions (PDF), 
  we  evaluate both the numerator and the
 denominator of the asymmetries  with
   NLO PDF\footnote{In  \cite{Bernreuther:2010ny} a different 
     procedure was used. 
    The numerators  were evaluated with 
    NLO PDF while in the  denominators LO PDF and the same value of
    $\alpha_s$ as in the numerator were used. This yields 
    slightly larger asymmetries  (by $\sim 4\%$) than those given below.}.
 
 We use $m_t=173.1$ GeV (on-shell mass), the QED coupling
   $\alpha(m_Z)=0.008$, and the weak mixing angle
   $\sin^2\theta_W=0.23$.  We use the  CTEQ6.6M
  PDF \cite{Nadolsky:2008zw} and the respective value of
  $\alpha_s(m_Z)$ provided by this set. The same value $\mu$ is used
  for the renormalization and the factorization scale, and numerical
  results are given for $\mu= m_t/2, m_t$, and $2 m_t$. 
  These scale choices are purely conventional.  
 The variation of the asymmetries within this range of $\mu$
   are no substitute for a realistic assessment of the theory uncertainites; see  the corresponding remarks on page \pageref{p8}
     below.\\

In Tables~\ref{tab:afbt-tev} and~\ref{tab:att-tev} we present our
results for the laboratory-frame and $\ttbar$
   rest-frame asymmetry \eqref{fobasy} and 
 \eqref{fobattasy}, respectively.  In the first rows,   
  the QCD and electroweak contributions to the numerators of these
  asymmetries are given. (Notice that contributions from quark flavors
  $q\neq u, d$ are, after convolution with the PDF, symmetric under
  interchange of $t$ and $\bar t$ and therefore do not contribute to
  the numerators.) In the row labeled $qg$  the sum of the
  contributions from the  $qg$ and ${\bar q} g$ fusion processes is
  given -- for the sole purpose of showing that it can be safely
  neglected for the Tevatron, which will be done in the following. 
  The tables show what has already been mentioned above: the  
 mixed QCD-QED and QCD-weak contributions increase the QCD asymmetries
 at the Tevatron by 18$\%$ and 5$\%$, respectively, i.e. in total by $23\%$.

In Table~\ref{tab:CDFd0newdat} we collect our results for  $A_{FB}^{t}$
 and for $A^{t\bar t}$ without and with cuts on $|\Delta
y|$ and  $M_{t\bar{t}}$ and list, for comparison, also  
   results from the  CDF and D$\emptyset$ experiments
  \cite{Aaltonen:2011kc,Abazov:2011rq,CDF2012}.
               The experimental results for the 
    asymmetries  are the 
  unfolded values at the $\ttbar$ production level, i.e.,  corrected for detector
 resolution  and acceptance.
The CDF results given in column 3 of this data
   are from lepton plus jets events based on an integrated luminosity of 8.7
  fb$^{-1}$, while column 2 resulted
   from the analysis of $L_{int}= 5.3$ fb$^{-1}$ \cite{Aaltonen:2011kc}.
 The D$\emptyset$ collaboration \cite{Abazov:2011rq}   did not find 
statistically  sensitive dependencies of  $A^{t\bar{t}}$ on $|\Delta y|$
and  $\mtt$ \cite{Abazov:2011rq} and therefore 
  did not  publish  unfolded numbers   for the
observables listed  in rows
2 - 5 of  Table~\ref{tab:CDFd0newdat}.  

\begin{center}
\begin{table}[pt]
 \centering{}\begin{tabular}{|c|c|c|c|c|}
\hline 
\multicolumn{1}{|c}{$N_{FB}^{t}$ (pb)} &  & $\mu={m_{t}}/{2}$  & $\mu=m_{t}$  & $\mu=2m_{t}$\tabularnewline
\hline 
$O(\alpha_{s}^{3})$  & $u\bar{u}$  & $0.3328$ & $0.2183$  &
$0.1489$ \tabularnewline  
\cline{3-5} 
 & $d\bar{d}$  & $0.0591$ & $0.0381$  & $0.0257$\tabularnewline 
\cline{3-5} 
 & $qg$ & $4.1 \times 10^{-5}$ & $2.6 \times 10^{-5}$ & $1.7 \times 10^{-5}$\tabularnewline
\hline 
$O(\alpha^{2})$  & $u\bar{u}$  & $9.4 \times 10^{-3}$ & $8.3 \times
10^{-3}$  & $7.4 \times 10^{-3}$\tabularnewline
\cline{3-5} 
 & $d\bar{d}$  & $1.2 \times 10^{-3}$ & $1.1 \times 10^{-3}$  & $9.2
 \times 10^{-4}$\tabularnewline
\hline 
$O(\alpha\alpha_{s}^{2})_{weak}$  & $u\bar{u}$  & $7.1 \times 10^{-3}$
 & $5.2 \times 10^{-3} $  &
$3.8 \times 10^{-3}$ \tabularnewline
\cline{3-5} 
 & $d\bar{d}$  & $-2.2  \times 10^{-3}$ & $-1.6 \times 10^{-3}$  & $-1.2 \times 10^{-3}$\tabularnewline
\hline 
$O(\alpha\alpha_{s}^{2})_{QED}$  & $u\bar{u}$  & $0.0692$ & $0.0502$  & $0.0375$\tabularnewline
\cline{3-5} 
 & $d\bar{d}$  & $-6.1 \times  10^{-3}$ & $-4.4 \times 10^{-3}$
 & $-3.2 \times 10^{-3}$\tabularnewline
\hline 
\multicolumn{1}{|c}{total} &  & $0.4701$ & $0.3151$  & $0.2198$\tabularnewline
\hline \hline
\multicolumn{1}{|c}{$\sigma_{QCD}^{LO}$ (pb)} &  & $7.618$ &
$5.456$ & $4.030$\tabularnewline
\hline \hline
\multicolumn{1}{|c}{$A_{FB}^{t}$ (\%)} &  & $6.17$ &
$5.77$ & $5.46$\tabularnewline
\hline
\end{tabular}
\caption{The contributions to the numerator of the  
 $t$-quark forward-backward laboratory-frame asymmetry \eqref{fobasy} at the
 Tevatron  for three different scales. The
 denominator of \eqref{fobasy}, $\sigma_{QCD}^{LO}=\sigma_{t{\bar t}}$, is computed at leading-order QCD. 
 }
\label{tab:afbt-tev}
\end{table}
\end{center}

\begin{center}
\begin{table}[pb]
 \centering{} \begin{tabular}{|c|c|c|c|c|}
\hline 
\multicolumn{1}{|c}{$N^{t\bar{t}}$~(pb)} &  & $\mu={m_{t}}/{2}$  & $\mu=m_{t}$  & $\mu=2m_{t}$\tabularnewline
\hline 
$O(\alpha_{s}^{3})$  & $u\bar{u}$  & $0.5014$ & $0.3297$  & 
   $0.2251$\tabularnewline
\cline{3-5} 
 & $d\bar{d}$  & $0.0899$ & $0.0582$  & $0.0392$\tabularnewline
\cline{3-5} 
 & $qg$ & $7.6 \times 10^{-5}$ & $3.4 \times 10^{-5}$ & $2.9 \times 10^{-5}$\tabularnewline
\hline 
$O(\alpha^{2})$  & $u\bar{u}$  & $1.47 \times 10^{-2}$ & $1.29 \times
10^{-2}$  & $1.15 \times 10^{-2}$\tabularnewline
\cline{3-5} 
 & $d\bar{d}$  & $1.9 \times 10^{-3}$ & $1.6 \times 10^{-3}$  & $1.5
 \times 10^{-3}$\tabularnewline
\hline 
$O(\alpha\alpha_{s}^{2})_{weak}$  & $u\bar{u}$  & $10.7 \times
10^{-3}$ & $7.8 \times 10^{-3}$  & $5.8 \times 10^{-3}$\tabularnewline
\cline{3-5} 
 & $d\bar{d}$  & $-3.4 \times 10^{-3}$ & $-2.4  \times 10^{-3}$  &
 $-1.8 \times 10^{-3}$\tabularnewline
\hline 
$O(\alpha\alpha_{s}^{2})_{QED}$  & $u\bar{u}$  & $0.1047$ & $0.0761$  & $0.0569$\tabularnewline
\cline{3-5} 
 & $d\bar{d}$  & $-9.4   \times 10^{-3}$ & $-6.7 \times 10^{-3}$  &
 $-4.9 \times 10^{-3}$\tabularnewline
\hline 
\multicolumn{1}{|c}{total} &  & $0.7104$ & $0.4772$  & $0.3332$\tabularnewline
\hline \hline
 \multicolumn{1}{|c}{$\sigma_{QCD}^{LO}$~(pb)} &  & $7.618$ &
$5.456$ & $4.030$\tabularnewline
\hline \hline
\multicolumn{1}{|c}{$A^{t\bar{t}}$~(\%)} &  & $9.33$ & $8.75$ & $8.27$\tabularnewline
\hline
\end{tabular}
 \caption{The  contributions to the numerator of the $t\bar t$ rest-frame
  asymmetry \eqref{fobattasy} at the Tevatron
  for three different scales. }
 \label{tab:att-tev}
\end{table}
\end{center}

\begin{center}
 \begin{table}  
 \centering{} \begin{tabular}{|c|c|c|c|c|}
 \hline 
 & CDF \cite{Aaltonen:2011kc} & CDF \cite{CDF2012}  & D$\emptyset$
 \cite{Abazov:2011rq}& SM (this work) \tabularnewline \hline
 $A_{FB}^{t}$ & $0.150\pm 0.055$&  & & $0.058\pm 0.004$ \tabularnewline
$A^{t\bar t}$ & $0.158 \pm 0.075$& $0.162 \pm 0.047$ & $ 0.196 \pm
0.065$ & $0.088\pm 0.006$  \tabularnewline
$A^{t\bar{t}}(|\Delta y|\leq 1)$ & $0.026 \pm  0.118$& $0.088\pm 0.047$  &  & $0.061^{+0.004}_{-0.003}$ \tabularnewline
 $A^{t\bar{t}}(|\Delta y|> 1)$ & $0.611\pm  0.256$ & $0.433\pm 0.109$ & & $0.206^{+0.011}_{-0.010}$ \tabularnewline
 $A^{t\bar t}(M_{t\bar t}\leq  450~{\rm GeV})$ & $-0.116 \pm
0.153$ & $0.078 \pm 0.054$ & & $0.062^{+0.004}_{-0.003}$ \tabularnewline
$A^{t\bar t}(M_{t\bar t} >  450~{\rm GeV})$ &  $0.475 \pm 0.114$ &
$0.296 \pm 0.067$ & &   $0.129^{+0.008}_{-0.006}$  \tabularnewline  \hline
 \end{tabular}
 \caption{Unfolded experimental results from CDF
   \cite{Aaltonen:2011kc,CDF2012}
  and D$\emptyset$   \cite{Abazov:2011rq} for the laboratory- and
   $\ttbar$ rest-frame  asymmetry at the
   Tevatron without and with cuts on $|\Delta y|$ and $M_{t\bar t}$
   and our SM predictions (errors are scale-uncertainties only).}
 \label{tab:CDFd0newdat}
 \end{table}
 \end{center}

 The D$\emptyset$  and recent CDF results \cite{CDF2012} on the inclusive
 rest-frame asymmetry $A^{t\bar{t}}$ are within  $\sim 1.5
\sigma$ of  our SM prediction.  As to the rest-frame
asymmetry with cuts: For $A^{t\bar{t}}(|\Delta y|\leq 1)$ and 
$A^{t\bar t}(M_{t\bar t}\leq  450~{\rm GeV})$, and 
   the recent CDF \cite{CDF2012} 
  and our SM results agree well, while
   the recent CDF determinations $A^{t\bar{t}}(|\Delta y|> 1)$ and
   of    $A^{t\bar t}(M_{t\bar t}>  450~{\rm GeV})$   
  deviates from our SM predictions  by $\sim
 2 \sigma$ and $\sim  2.4 \sigma$, respectively.

The asymmetry  $A^{t\bar t}$ 
 increases approximately linearly with $|\Delta y|$ and 
$M_{t\bar{t}}$. The slopes of these straight lines that were recently determined
   by the CDF experiment \cite{CDF2012} are 
  significantly larger than  those obtained in the SM. A cut on the
 transverse momentum of the $\ttbar$ system has a significant effect
  on the size of the charge asymmetries. For instance, 
 selecting $\ttbar$ events with
  low $p_\perp^{\ttbar}$ significantly increases the 
  asymmetries \cite{Kuhn:2011ri}. This is due to the fact that 
 the positive inclusive NLO QCD asymmetries  are generated by the
           contribution from Born times virtual and soft gluon terms,
           which is positive and the contribution from hard gluon radiation,
          which is negative. 

Experiments usually compare their results with predictions made with
  one of the widely used NLO QCD Monte-Carlo generators
  \cite{Frixione:2008ym,Frixione:2007nw,MCFM,Campbell:2010ff}. 
 In these programs the electroweak
   contributions to the asymmetries are not included.
  More importantly, in these Monte-Carlo calculations
   the denominators of the
asymmetries  are determined with NLO QCD parton matrix elements, 
  which reduces the asymmetry
by up to  $\sim 30\%$ as compared to the procedure employed by us and in 
\cite{Hollik:2011ps,Kuhn:2011ri}.

We now compare our results with 
  other recent SM calculations of  $A_{FB}^{t}$ and
$A^{t\bar{t}}$. In \cite{Hollik:2011ps}  the
various contributions to the asymmetries were also given in detail. 
  Although we use a different PDF set 
than  \cite{Hollik:2011ps}, our
results of Tables~\ref{tab:afbt-tev}, 
\ref{tab:att-tev} and \ref{tab:CDFd0newdat} 
agree well with the numbers of the corresponding Tables of that
reference. Ref. \cite{Kuhn:2011ri} also used  a PDF set different from
ours and employed the strategy of evaluating  the numerators and the
denominators of the asymmetries  with LO PDF. Moreover,
 the mixed QCD-weak corrections, which make only a small contribution,
 were taken into account only approximately in  \cite{Kuhn:2011ri}. 
  Our results agree also with those of  \cite{Kuhn:2011ri}.
 The recent fixed-order NLO QCD computation of $A_{FB}^{t}$ of
  \cite{Campbell:2012uf}
   uses NLO matrix elements in the denominator and therefore gets
    a smaller value than our pure QCD result $7.1(6)\%$ 
   (cf. Table~\ref{tab:att-tev}). Moreover, with this procedure
   the uncertainties due
   to scale variations become significantly larger than those given in 
Table~\ref{tab:CDFd0newdat}.
 Ref.~\cite{Ahrens:2011uf}  computed the above asymmetries in pure
 QCD, at NLO plus next-to-next-to-leading logarithmic accuracy (NLO +
 NNLL), by performing corresponding resummations of logarithms due
 to soft and collinear gluons. 
\label{p8}
One expects that these resummations
 (cf. also  \cite{Almeida:2008ug,Kidonakis:2011zn})
 provide more realistic estimates  of the scale uncertainties
 than those resulting from  the  fixed order  NLO predictions.
  The central values of  $A_{FB}^{t}$ and
$A^{t\bar{t}}$, without and with the above cuts on the  latter
asymmetry,  given in~\cite{Ahrens:2011uf} are essentially the same as
those obtained at fixed order NLO QCD. This may not be surprising
because soft and collinear radiation from top quarks (which is the
physics behind  taking into account threshold resummations)
 does not change the directions of $t$ and $\bar t$ and thus the 
 asymmetries in an essential way. 
 
In concluding this section we recall that the recent CDF determination of the
   high-mass asymmetry has reduced, but not erased the tension with the 
 existing NLO SM predictions.
  We emphasize that the uncertainties due to scale 
  variations of our SM results given in Tables~\ref{tab:afbt-tev}, 
\ref{tab:att-tev} and \ref{tab:CDFd0newdat} underestimate the theory
errors\footnote{According to \cite{Brodsky:2012ik} a judicious choice of scale-setting leads
  to a significant increase of the QCD-induced asymmetries.},
 which are, more realistically, of the order of $\sim 30\%$. It remains to be 
  seen whether a complete fixed-order NNLO QCD computation of the asymmetries
  will alleviate this tension. 

\subsection{LHC}
\label{suse:lhc}

Let us first recall the salient features of the 
   charge asymmetries in
  top-quark pair production in $p p$ collisions, $p p \to \ttbar + X$.
   At the LHC, the
 initial $pp$ state  is an eigenstate of parity.
 Thus, $A_{FB}^t=A_{FB}^{\bar t}=0$ 
 as long as only parity-invariant interactions  are taken into account.
 In fact, the parity-violating terms of the weak corrections
  appear  only in the $t$- and/or $\bar t$-spin dependent
 terms of the inclusive partonic $\ttbar$ 
  production density matrices and do, therefore,
 not contribute to the inclusive $\ttbar$ asymmetries
   when making predictions for top quarks summed over
 their spins.  As a consequence, at the LHC the 
  differential charge asymmetry  $A(y)$ induced  
  by the SM interactions must be symmetric with respect to $y=0$. 
  However, QCD predicts that for large values of $|y|$ of the (anti)top
  rapidity, the $\ttbar$ sample is such that there  are more $t$ than
  $\bar t$ quarks,
 while for small values of $|y|$ it is the other way around.
  Therefore, in the SM 
        the differential charge 
   asymmetry $A(y)>0$ in the forward and backward regions, while  
  $A(y)<0$ in the central region.  Thus one can define non-zero
  (integrated) asymmetries. The dominant contributions to the numerator
   of $A(y)$ are again due to the
   antisymmetric part ($t\leftrightarrow {\bar t}$) of
    the $q\bar q$ differential cross section. Contrary to the Tevatron,
     the  antisymmetric contributions from $qg$ fusion are not 
     negligibly small at the LHC (see below). 

Because at the LHC  the fraction of $\ttbar$ production by $q\bar q$ annihilation is significantly smaller
    than by $gg$ fusion, it
   is clear that the charge asymmetries are  smaller
   than at the Tevatron.  With suitable cuts one
   may enhance the asymmetries. For instance, in the SM one 
  expects that  the charge asymmetries 
   increase in magnitude with $\mtt$ because at the LHC the $q\bar q$ luminosity
  increases with respect to the $gg$ luminosity for increasing pair-invariant
   mass. Other ways to enhance the  $q\bar q$-initiated fraction
   of $\ttbar$ and thus the ratio of the antisymmetric and
   symmetric part of the $\ttbar$
   cross section is to select forward and/or backward events, to
   select $\ttbar$ events whose c.m. frame is highly boosted along the
   beam axis with  respect to the laboratory frame, or to put
  a cut on the transverse momentum of the $\ttbar$ system.
    These observations
   have led to a number of suggestions for LHC observables
   \cite{Kuhn:1998kw,Antunano:2007da,Wang:2010du,Chatrchyan:2011hk,Aad:2012ug,Xiao:2011kp,Hewett:2011wz,Arguin:2011xm,AguilarSaavedra:2011cp,Kuhn:2011ri,Alvarez:2012vq,Alvarez:2012uh} that exhibit small, but non-zero SM-induced charge asymmetries and are useful
 in discriminating between various new physics models which were
 proposed to explain the Tevatron asymmetry.

In the following analysis of various LHC charge
   asymmetries, we have taken into account in the computation
   of the respective numerators 
  the ${\cal O}(\alpha_s^3)$ QCD
 and the ${\cal O}(\alpha^2)$ and  ${\cal
   O}(\alpha_s^2\alpha)$ electroweak contributions as outlined
  in Sect.~\ref{suse:tevatron}. As mentioned above, the
    antisymmetric contributions from $qg$ fusion of
    $O(\alpha_{s}^{3})$ 
 are not negligible at the LHC. For completeness, we take into account
 also the mixed QCD-QED corrections of $O(\alpha\alpha_{s}^{2})$ to 
 $qg$ fusion -- see below. 
 The denominators of the asymmetries
  are evaluated again with LO QCD matrix elements and the NLO PDF  set CTEQ6.6M.

\subsubsection*{Central and edge charge asymmetry}

Choosing a cut $y_c$ on the rapidities
  of the $t$ and $\bar t$ quarks, one
    may define central and edge (or forward)
    charge asymmetries
  $A_{C}$, $A_{E}$  \cite{Antunano:2007da,Xiao:2011kp,Hewett:2011wz}:
\begin{equation} \label{eq:centcasy}
A_{C}(y_c) =\frac{N(|y_{t}|\leq
  y_{c})-N(|y_{\bar{t}}|\leq
  y_{c})}{N(|y_{t}|\leq
  y_{c})+ N(|y_{\bar{t}}|\leq y_{c})} \, , 
\end{equation}
\begin{equation} \label{edgeasy}
A_{E}(y_{c})=\frac{N(y_{c}\leq|y_{t}|)-
 N(y_{c}\leq|y_{\bar{t}}|)}{N(y_{c}\leq|y_{t}|)
 +N(y_{c}\leq|y_{\bar{t}}|)} \, ,
\end{equation}
where the (anti)top rapidities are defined in the laboratory frame.
  The above discussion tells us that for suitably chosen $y_c$, the
   central asymmetry $A_{C}(y_c)<0$  and $A_{E}(y_{c})>0$ in the SM.
 Because the fraction of $q \bar q$ initiated $\ttbar$ events,
 $\sigma_{q{\bar q} \to \ttbar}/\sigma_{\ttbar}$, is
 enhanced in the forward/backward region, $A_{E}$ will in general be larger
 than $|A_{C}|$. On the other hand, the event numbers decrease rapidly
 with  increasing $|y|$; i.e., $y_c$ must be chosen appropriately 
  for each of these observables in order to optimize the  statistical
  sensitivity of $A_{E}$.

For the computation of the  central asymmetry we 
 choose $y_{c}=1$  and take into account $\ttbar$ events with
   $M_{tt}\geq M_{c}$. We choose $M_c = 2m_t$,
    0.5 TeV, 0.7 TeV and 1 TeV. The various contributions
 to the numerator and the resulting values of $A_{C}(y_c=1)$ 
   at 7 TeV center-of-mass energy are given in
   Table~\ref{tab:AC1-lhc}. 
   The size of the  $O(\alpha\alpha_{s}^{2})$ mixed QCD-QED corrections to
  $q\bar q$ initiated contributions relative to  those of
  $O(\alpha_{s}^{3})$ QCD is now $\sim 13 \%$,  which,   as already
  mentioned in Sect.~\ref{suse:tevatron}, is due to the fact that
  the ratio of $u\bar u$ versus $d\bar d$ annihilation is 2:1 at the LHC as compared
   to $4:1$ for $p \bar p$ collisions.
   The size of the 
    $O(\alpha_{s}^{3})$ QCD contributions from $qg$ fusion amount 
   to about $5\%$ ($M_c =
    2m_t$) of   the $q\bar q$ contributions.
 At $\sqrt{s}=14$ TeV and  $M_c = 1$ TeV, they rise to $\sim
 17\%$. Here, and also for all other LHC asymmetries discussed below, 
  we take into account also the mixed QCD-QED corrections of $O(\alpha\alpha_{s}^{2})$ to 
 $qg\to \ttbar q$ which are of the same order of magnitude as the
 mixed QCD-weak corrections of $O(\alpha\alpha_{s}^{2})$,  as shown in
 Table~\ref{tab:AC1-lhc}. The size of these corrections can be easily
 understood. By diagram inspection at the level of initial partons
   one obtains that the ratio 
 $f_q = O(\alpha\alpha_{s}^{2})_{QED}/O(\alpha_{s}^{3})$ for  $qg\to
   \ttbar q$ is given by 
 \begin{equation} \label{fqedr}
  f_q = \frac{4\alpha Q_q Q_t}{\alpha_s d^2_{abc}/4} = \frac{24
    \alpha Q_q Q_t}{5 \alpha_s} \, ,
 \end{equation}
 where  $d^2_{abc}=40/3$. For $p p$ collisions at the LHC one gets
 therefore the ratio
 \begin{equation} \label{fqedlhc}
 f^{QED} = \frac{4 f_u + 2 f_d}{6} = \frac{16 \alpha}{15 \alpha_s} \, .
  \end{equation}
Using $\alpha_s\simeq 0.11$ and $\alpha\simeq 0.008$, one gets 
 $f^{QED}\simeq 0.078$. This estimate explains the respective results
 of Table~\ref{tab:AC1-lhc} which were obtained by integrating the 
    respective matrix elements and PDF.

   In Table~\ref{tab:AC3-lhc} the values of $A_{C}(y_c=1)$  are given for  $\sqrt{s}=7, 8,$ and $14$ TeV, 
    both for QCD and for  QCD plus electroweak contributions.
    The given uncertainties are due to scale variations. As above we
    choose $\mu=m_t/2, m_t,$ and $2m_t$.  
   The asymmetry $A_{C}(y_c=1)$ increases with
   increasing  lower bound  $M_c$ on $\mtt$. But, as  the numbers
   for the denominator $D_C$ in  Table~\ref{tab:AC1-lhc} show, 
   the  event numbers decrease rapidly with increasing 
   $M_c$. The ratio of the electroweak and QCD contributions to $A_{C}(y_c=1)$ 
    is $13\%$ for $\sqrt{s}=7$ TeV (no cut on $\mtt$) and increases slightly
    to $16\%$ for $\sqrt{s}=14$ TeV and $\mtt>1$ TeV.

The various contributions to the numerator of the edge asymmetry are
   collected in Table~\ref{tab:EC1-lhc} for $\sqrt{s}=7$ TeV,
  and   $A_{E}(y_c)$ is given as a function of $y_c$ for $\sqrt{s}=7, 8$ and $14$ TeV
   in Table~\ref{tab:EC3-lhc}.
 The statistical significances of  $A_{E}$ and $A_C$ are  of
comparable size. For instance, $A_{E}(y_c=1)=1\%$ while $A_{C}(y_c=1)=-0.6\%$
  at 7 TeV. The smaller value of $A_C$ is compensated by the larger
  number of  events with $y_c \leq 1$.

\begin{center}
\begin{table}[h]
\centering{} \begin{tabular}{|c|c|c|c|c|c|}
\hline 
\multicolumn{2}{|c|}{$M_{c}$} &  $2m_t$ & 0.5 TeV  & 0.7 TeV  & 1 TeV\tabularnewline
\cline{1-2} 
\multicolumn{2}{|c|}{$N_{C}$(pb)} &  &  &  & \tabularnewline
\hline
$O(\alpha_{s}^{3})$  & $q\bar{q}$  & $-0.6270$ & $-0.3718$ & $-0.1202$ & $-2.274\times10^{-2}$\tabularnewline
\cline{2-6} 
 & $qg$ & $-0.0379$ & $-0.0227$ & $-0.0100$ & $-0.0020$\tabularnewline
\hline 
$O(\alpha^{2})_{weak}$  & $q\bar{q}$  & $-0.0234$ & $-0.0134$ & $-0.0040$ & $-6\times10^{-4}$\tabularnewline
\hline 
$O(\alpha\alpha_{s}^{2})_{weak}$  & $q\bar{q}$  & $-2.5\times10^{-3}$ & $-1.3\times10^{-3}$ & $-4.5\times10^{-4}$ & $-9\times10^{-5}$\tabularnewline
\cline{2-6} 
 & $qg$ & $7.1\times10^{-3}$ & $1.7\times10^{-3}$ & $-4.4\times10^{-4}$ & $-2.1\times10^{-4}$\tabularnewline
\hline 
$O(\alpha\alpha_{s}^{2})_{QED}$  & $q\bar{q}$  & $-7.85\times10^{-2}$ & $-4.81\times10^{-2}$ & $-1.53\times10^{-2}$ & $-2.7\times10^{-3}$\tabularnewline
\cline{2-6} 
 & $qg$ & $-2.7\times10^{-3}$ & $-2.0\times10^{-3}$ & $-7\times10^{-4}$ & $-1\times10^{-4}$\tabularnewline
\hline 
\multicolumn{2}{|c|}{Total} & $-0.7648$ & $-0.4576$ & $-0.1512$ & $-0.0286$\tabularnewline
\hline
\hline 
\multicolumn{2}{|c|}{$D_{QCD}^{LO}$(pb)} & $126.76$ & $45.76$ & $9.89$ & $1.35$\tabularnewline
\hline
\hline 
\multicolumn{2}{|c|}{$A_{C}$(\%)} & $-0.60$ & $-1.00$ & $-1.53$ & $-2.13$\tabularnewline
\hline
\end{tabular}
\caption{The contributions to the numerator and denominator of  $A_{C}(y_c=1)$,
  defined  in   \eqref{eq:centcasy}, for
  $\mu=m_{t}$ at the LHC (7 TeV).}
\label{tab:AC1-lhc}
\end{table}

\par\end{center}

\begin{center}
\begin{table}[pt]
\centering{} \begin{tabular}{|c|c c|c|c|c|c|}
\hline
$\sqrt{s}$ &  & &$M_c = 2 m_t$ &  0.5 TeV  & 0.7 TeV  & 1 TeV  \tabularnewline
\hline 
 7 TeV  & QCD: &$A_{C}$~(\%) &$-0.53~(3)$ & $-0.86~(3)$ & $-1.32~(5)$ & $-1.77~(7)$ \tabularnewline 
 & QCD + EW:  &$A_{C}$~(\%) & $-0.60~(3)$ & $-1.00~(4)$ & $-1.53~(5)$ & $-2.07~(7)$ \tabularnewline \hline
 8 TeV  & QCD:      &$A_{C}$~(\%)  &$-0.47~(2)$ & $-0.76~(2)$ &
 $-1.18~(4)$ & $-1.66~(5)$  \tabularnewline 
        & QCD + EW: & $A_{C}$~(\%) &$-0.54~(3)$ & $-0.88~(4)$&
        $-1.37~(4)$& $-1.94~(5)$  \tabularnewline \hline\hline
       &  & & $M_c = 2 m_t$ & 0.5 TeV & 1 TeV & 2 TeV \tabularnewline \hline  
14 TeV  & QCD:      &$A_{C}$~(\%)  & $-0.26~(2)$ & $-0.45~(2)$ &$-1.09~(4)$ &
  $-1.90~(6)$ \tabularnewline
        & QCD + EW: & $A_{C}$~(\%) & $-0.30~(3)$ & $-0.52~(4)$ &  $-1.29~(5)$ &$-2.21~(5)$
 \tabularnewline \hline
\end{tabular}
\caption{The central charge asymmetry 
    $A_{C}(y_c=1)$  for  the LHC at  $7, 8,$ and $14$ TeV, for events
    with $\mtt\geq M_c$.
   The uncertainties are due to scale variations.
  }
\label{tab:AC3-lhc}
\end{table}
\end{center}

\begin{center}
\begin{table}[h]
\centering{} \begin{tabular}{|c|c|c|c|c|}
\hline 
\multicolumn{2}{|c|}{$N_{E}$ (pb)} & $Y_{C}=0.5$  & $Y_{C}=1$  & $Y_{C}=2$\tabularnewline
\hline
$O(\alpha_{s}^{3})$  & $q\bar{q}$  & $0.4325$ & $0.6270$ & $0.3117$\tabularnewline
\cline{2-5} 
 & $qg$ & $0.0238$ & $0.0379$ & $0.0125$\tabularnewline
\hline 
$O(\alpha^{2})_{weak}$  & $q\bar{q}$  & $0.0154$ & $0.0234$ & $0.0103$\tabularnewline
\hline 
$O(\alpha\alpha_{s}^{2})_{weak}$  & $q\bar{q}$  & $1.2\times10^{-3}$ & $2.6\times10^{-3}$ & $2.1\times10^{-3}$\tabularnewline
\cline{2-5} 
 & $qg$ & $-4.6\times10^{-3}$ & $-7.1\times10^{-3}$ & $-4.1\times10^{-3}$\tabularnewline
\hline 
$O(\alpha\alpha_{s}^{2})_{QED}$  & $q\bar{q}$  & $0.0488$ & $0.0785$ & $0.0450$\tabularnewline
\cline{2-5} 
 & $qg$ & $1.8\times10^{-3}$ & $2.7\times10^{-3}$ & $1.1\times10^{-3}$\tabularnewline
\hline 
\multicolumn{2}{|c|}{Total} & $0.5189$ & $0.7648$ & $0.3776$\tabularnewline
\hline 
\multicolumn{2}{|c|}{$D_{QCD}^{LO}$(pb)} & $131.86$ & $73.77$ & $10.27$\tabularnewline
\hline 
\multicolumn{2}{|c|}{$A_{E}$(\%)} & $0.39$ & $1.04$ & $3.69$\tabularnewline
\hline
\end{tabular}\caption{The contributions to the numerator and
  denominator of $A_{E}$,
  defined in \eqref{edgeasy},  for $\mu=m_{t}$ at
  the LHC (7 TeV).}
\label{tab:EC1-lhc}
\end{table}

\par\end{center}

The center and edge asymmetries were computed before in \cite{Kuhn:2011ri}
 at NLO QCD including electroweak corrections, as functions of $y_c$
 for 7 and 14 TeV.  Ref.  \cite{Kuhn:2011ri} evaluated the numerators and
 denominators of the asymmetries with the PDF set \cite{Martin:2009iq} and
 took the purely weak corrections  only approximately into account. Our results above
 agree\footnote{The definition of the central  asymmetry $A_{C}$   in \eqref{eq:centcasy}   differs
    by a sign from that of  \cite{Kuhn:2011ri}.}, within the 
    given uncertainties, with  \cite{Kuhn:2011ri}.

\begin{center}
\begin{table}[pt]
\centering{} \begin{tabular}{|c|c c|c|c|c|}
\hline 
$\sqrt{s}$ &  & & $y_{c}=0.5$  & $y_{c}=1$  & $y_{c}=2$\tabularnewline
\hline 
   7 TeV & QCD: & $A_{E}$~(\%) & $0.35~(1)$ & $0.90~(3)$ & $3.16~(6)$\tabularnewline
 & QCD + EW: & $A_{E}$~(\%) & $0.39~(2)$ & $1.04~(4)$ & $3.69~(7)$\tabularnewline
\hline 
8 TeV & QCD: & $A_{E}$~(\%) & $0.29~(1)$ & $0.74~(3)$ & $2.69~(6)$   \tabularnewline
 &  QCD + EW: & $A_{E}$~(\%) &$0.31~(2)$  & $0.86~(3)$& $3.24~(6)$ \tabularnewline \hline 
14 TeV & QCD: & $A_{E}$~(\%) & $0.12~(1)$ & $0.32~(1)$ & $1.28~(5)$ \tabularnewline
 &  QCD + EW: & $A_{E}$~(\%) & $0.14~(1)$ & $0.37~(3)$ & $1.49~(9)$\tabularnewline
\hline
\end{tabular}
\caption{The edge asymmetry $A_{E}$ as a function
  of $y_c$ for the LHC at $7, 8,$ and $14$ TeV.  The uncertainties are due to scale variations.
}
\label{tab:EC3-lhc}
\end{table}
\end{center}

\subsubsection*{Cut-independent charge asymmetries}
 The CMS  \cite{Chatrchyan:2011hk,CMSchargeas} and ATLAS
 \cite{Aad:2012ug} experiments measured the following rapidity-cut independent charge asymmetries:
\begin{equation} \label{casyylhc}
A_{C}^{\Delta|y|}=\frac{N(\Delta|y|>0)-
 N(\Delta|y|<0)}{N(\Delta|y|>0)+
 N(\Delta|y|<0)} \, ,
\end{equation}

\begin{equation} \label{casyetalhc}
A_{C}^{\Delta|\eta|}=\frac{N(\Delta|\eta|>0)-
 N(\Delta|\eta|<0)}{N(\Delta|\eta|>0)+
 N(\Delta|\eta|<0)} \, ,
\end{equation}
where $\Delta|y|=|y_{t}|-|y_{\bar{t}}|$ and likewise for the
  pseudorapidities,  $\Delta|\eta|=|\eta_{t}|-|\eta_{\bar{t}}|$,
  in the laboratory frame. \\
 We compute these asymmetries for $\ttbar$ events with
   $M_{tt}\geq M_{c}$. As above, we choose $M_c = 2m_t$ (i.e., all events),
    0.5 TeV, 0.7 TeV and 1 TeV. Our NLO QCD predictions  and those
    including the electroweak corrections are given
    in Tables~\ref{tab:casyy-lhc} and~\ref{tab:casyeta-lhc}
   for  $\sqrt{s}=7, 8,$ and  $14$ TeV. With a cut $M_{tt}\geq 1$ TeV, 
   the asymmetries $A_{C}^{\Delta|y|}$, $A_{C}^{\Delta|\eta|}$
    increase by a factor of about two.  The ratio of electroweak and QCD
    contributions to the asymmetries is $15\%$ for $\sqrt{s}=7$ TeV and no cut
    on $\mtt$, and it increases to $\gtrsim 20\%$ 
 at  $\sqrt{s}=14$ TeV and large $\mtt$.

The asymmetries $A_{C}^{\Delta|y|}$ and $A_{C}^{\Delta|\eta|}$ were
computed also in  \cite{Kuhn:2011ri} in the SM without a cut on $\mtt$. The
respective numbers in Tables~\ref{tab:casyy-lhc} and~\ref{tab:casyeta-lhc} agree with these results.

The experimental results of the CMS and ATLAS collaborations  are given in
Table~\ref{tab:cmsatl-lhc}. The results agree, within the present
uncertainties, with the SM predictions given 
 above\footnote{In view of the positive charge
  asymmetry measured at the Tevatron one expects the LHC 
  asymmetry $A_{C}$ to be
  positive, too, within the SM. However, there are examples
   of new physics models which yield a negative LHC asymmetry;
   see, e.g., \cite{Drobnak:2012cz}.}.

The recent CMS analysis \cite{CMSchargeas}, based on a data sample of
$L_{int}=4.7$ fb$^{-1}$, measured the charge asymmetry  $A_{C}^{\Delta
  |y|}$ also differentially; in particular as a function of $\mtt$. 
 The respective data given in  \cite{CMSchargeas} agree, within the
 still large experimental errors, with
 our SM prediction of the $\mtt$ dependence of $A_{C}^{\Delta |y|}$
 given in Table~\ref{tab:casyy-lhc}.

\begin{center}
\begin{table}[pt]
\centering{} \begin{tabular}{|c|c c|c|c|c|c|}
\hline 
$\sqrt{s}$ & & & $M_c = 2 m_t$ & 0.5 TeV  &  0.7 TeV  &  1 TeV  \tabularnewline
\hline 
7 TeV &  QCD:& $A_{C}^{\Delta |y|}$~(\%) & $1.07~(4)$ & $1.27~(4)$ & $1.68~(4)$ & $2.06~(5)$  \tabularnewline
 &  QCD + EW: & $A_{C}^{\Delta |y|}$~(\%) & $1.23~(5)$ & $1.48~(4)$ & $1.95~(4)$ & $2.40~(6)$  \tabularnewline
\hline 
8 TeV & QCD:  & $A_{C}^{\Delta |y|}$~(\%) &$0.96~(4)$ &$1.14~(4)$ &$1.48~(4)$  & $1.85~(4)$   \tabularnewline
 &  QCD + EW: & $A_{C}^{\Delta |y|}$~(\%) & $1.11~(4)$ & $1.33~(5)$ &
 $1.73~(5)$ & $2.20~(5)$  \tabularnewline
\hline \hline
  & & & $M_c = 2 m_t$ & 0.5 TeV  &  1 TeV  &  2 TeV  \tabularnewline \hline
14 TeV &  QCD: & $A_{C}^{\Delta |y|}$~(\%) & $0.58~(3)$ & $0.74~(3)$ 
 & $1.11~(5)$ & $1.72~(10)$ \tabularnewline
 &  QCD + EW: &  $A_{C}^{\Delta |y|}$~(\%) & $0.67~(4)$ &  $0.86~(5)$ &   $1.32~(8)$& $2.12~(10)$ \tabularnewline
\hline
\end{tabular}
\caption{The charge asymmetry $A_{C}^{\Delta |y|}$ defined in 
  \eqref{casyylhc} at the LHC,  for $M_{t\bar{t}} \geq M_{c}$. 
}
\label{tab:casyy-lhc}
\end{table}
\end{center}

\begin{center}
\begin{table}[h]
\centering{} \begin{tabular}{|c|c c|c|c|c|c|}
\hline 
$\sqrt{s}$ &  & & $M_{c}=2 m_t$ & 0.5 TeV  & 0.7 TeV  & 1 TeV   \tabularnewline
\hline 
7 TeV &  QCD:& $A_{C}^{\Delta |\eta|}$~(\%) & $1.36~(6)$ & $1.39~(5)$ & $1.76~(5)$ & $2.15~(5)$   \tabularnewline
 & QCD + EW: &  $A_{C}^{\Delta|\eta|}$~(\%) & $1.56~(7)$ & $1.64~(6)$ & $2.06~(5)$ & $2.52~(5)$   \tabularnewline
\hline 
8 TeV &  QCD:& $A_{C}^{\Delta|\eta|}$~(\%) &$1.24~(6)$  & $1.25~(5)$ &
   $1.56~(4)$&   $1.93~(4)$  \tabularnewline 
 & QCD + EW: &  $A_{C}^{\Delta|\eta|}$~(\%) &$1.43~(7)$  & $1.47~(5)$ & $1.84~(5)$ & $2.30~(5)$  \tabularnewline
\hline \hline
   &  & & $M_{c}=2 m_t$ & 0.5 TeV  & 1 TeV  & 2 TeV   \tabularnewline\hline
14 TeV &  QCD:& $A_{C}^{\Delta|\eta|}$~(\%) & $0.83~(5)$ & $0.84~(5)$ &  $1.16~(5)$ & 1.82~(7) \tabularnewline 
 & QCD + EW: & $A_{C}^{\Delta|\eta|}$~(\%) & $0.96~(6)$ & $1.00~(6)$ &   $1.44~(6)$ & $2.38~(8)$ \tabularnewline
\hline
\end{tabular}
\caption{The charge asymmetry $A_{C}^{\Delta|\eta|}$ defined in  \eqref{casyetalhc}
 at the LHC,  for $M_{t\bar{t}} \geq M_{c}$.
}
\label{tab:casyeta-lhc}
\end{table}
\end{center}

\begin{center}
\begin{table}[h]
\centering{} \begin{tabular}{|c|c|c|}
\hline 
 & $A_{C}^{\Delta |y|}$~(\%) &  $A_{C}^{\Delta|\eta|}$~(\%) \tabularnewline
 \hline
 CMS   &     $0.4 \pm 1.0 \pm 1.2$     \cite{CMSchargeas}      & $-1.7
 \pm 3.2^{+2.5}_{-3.6}$  \cite{Chatrchyan:2011hk} 
\tabularnewline  \hline
 ATLAS      &  $-1.8 \pm 2.8 \pm 2.3$ \cite{Aad:2012ug}  &         \tabularnewline \hline                                   
\end{tabular}
\caption{
 CMS \cite{Chatrchyan:2011hk,CMSchargeas} and ATLAS
  \cite{Aad:2012ug} results at the LHC (7 TeV).
}
\label{tab:cmsatl-lhc}
\end{table}
\end{center}

\subsubsection*{Boosted charge asymmetry}
Another way to enhance the $\ttbar$ charge asymmetries at the LHC is to
select $\ttbar$ events whose center-of-mass frame has a considerable
 Lorentz boost with respect to the beam axis. The velocity 
 of the $\ttbar$ system along the beam axis is given by 
\begin{equation}\label{betboost}
 \beta= \frac{|p_t^z + p_{\bar t}^z|}{E_t + E_{\bar t}} \, ,
\end{equation}
where $p^z$ and $E$ is the corresponding longitudinal momentum
component and energy in the laboratory frame, respectively. 
Ref. \cite{AguilarSaavedra:2011cp} proposed to evaluate the asymmetries
    \eqref{casyylhc}, \eqref{casyetalhc} for $\ttbar$ events with 
$\beta$ larger than a certain minimal value $\beta_{min}$. By
increasing $\beta_{min}$ the $q \bar q$ initiated $\ttbar$ sample and
therefore the $\ttbar$ charge asymmetry grows. 

This `boosted charge asymmetry' is similar to the one-sided charge
asymmetry  \cite{Wang:2010du} that will be discussed below. 
 Using the variable $\beta$ rather than the  longitudinal momentum
  $|p_t^z + p_{\bar t}^z|$ has obvious experimental advantages: the
  ratio $\beta$ is less affected by uncertainties due to jet energy
  scale and resolution. 

Table~\ref{tab:booscas-lhc} contains our results for the
 charge  asymmetry \eqref{casyetalhc}    as a function
 of $\beta_{min}$, i.e. $A_{C}^{\Delta|\eta|}(\beta_{min})$,
 for the LHC at 7, 8, and 14 TeV. 
 As expected the SM-induced asymmetry increases  with
increasing  $\beta_{min}$ -- of course, again at the expense of
decreasing $\ttbar$ samples.
 The contribution of the electroweak interactions is about $15 \%$
 compared to the pure QCD asymmetry.

\begin{table}
\begin{centering}
\begin{tabular}{|c|c|c|c|c|}
\hline 
\multicolumn{2}{|c|}{$A_{C}^{\Delta|\eta|}$ (\%) } & 7 TeV & 8 TeV & 14 TeV\tabularnewline
\hline
\hline
$\beta_{min}=0.1$ & QCD & 1.41~(7) & 1.28~(5) & 0.86~(5)\tabularnewline 
 & QCD+EW               & 1.62~(7) & 1.48~(7) & 1.00~(7)\tabularnewline
\hline
$\beta_{min}=0.2$ & QCD  & 1.50~(7) & 1.37~(6) & 0.90~(5)\tabularnewline
 & QCD+EW                & 1.72~(8) & 1.57~(7) & 1.04~(7)\tabularnewline
\hline
$\beta_{min}=0.3$ & QCD  & 1.63~(7) & 1.47~(6) & 0.95~(5)\tabularnewline
               & QCD+EW  & 1.87~(8) & 1.69~(8) & 1.10~(7)\tabularnewline
\hline
$\beta_{min}=0.4$ & QCD  & 1.77~(8) & 1.56~(7) & 1.01~(6)\tabularnewline 
 & QCD+EW                & 2.02~(9) & 1.79~(8) & 1.17~(8)\tabularnewline
\hline
$\beta_{min}=0.5$ & QCD  & 1.87~(10) & 1.69~(7) & 1.10~(6)\tabularnewline
 & QCD+EW                & 2.16~(10) & 1.95~(9) & 1.27~(8)\tabularnewline
\hline
$\beta_{min}=0.6$ & QCD  & 2.07~(10) & 1.86~(8) & 1.21~(6)\tabularnewline
 & QCD+EW                & 2.38~(10) & 2.14~(10) & 1.39~(9)\tabularnewline
\hline
$\beta_{min}=0.7$ & QCD  & 2.30~(10) & 2.08~(8) & 1.33~(7)\tabularnewline
 & QCD+EW                & 2.65~(11) & 2.40~(11) & 1.53~(10)\tabularnewline
\hline
$\beta_{min}=0.8$ & QCD: & 2.67~(12) & 2.39~(10) & 1.54~(9)\tabularnewline
 & QCD+EW                & 3.08~(12) & 2.76~(12) & 1.78~(11)\tabularnewline
\hline
$\beta_{min}=0.9$ & QCD  & 3.22~(13) & 2.95~(12) & 1.90~(10)\tabularnewline
 & QCD+EW                & 3.74~(12) & 3.42~(13) & 2.20~(12)\tabularnewline
\hline
\end{tabular}
\par\end{centering}
\caption{The charge asymmetry $A_{C}^{\Delta|\eta|}$ defined in
   \eqref{casyetalhc}  for $\ttbar$ events with
  $\beta={|p_{t}^{z}+p_{\bar{t}}^{z}|}/({E_{t}+E_{\bar{t}}})>\beta_{min}$
  at the LHC. }
\label{tab:booscas-lhc}
\end{table}

\subsubsection*{One-sided charge asymmetry}
Finally, we consider an asymmetry introduced and computed within QCD  in
 \cite{Wang:2010du}. Let $P^z_{\ttbar}$ be the component along the
 beam of the sum of the $t$ and $\bar t$ momenta, ${\bf P}_{\ttbar}
 ={\bf p}_t + {\bf p}_{\bar t}$,
   in the laboratory
 frame. For $ p p$ collisions,  a non-zero charge asymmetry may be obtained 
 by selecting $\ttbar$ events with  $P^z_{\ttbar}>0$,  or 
 events with  $P^z_{\ttbar}<0$. One may define a one-sided charge asymmetry by
 \begin{equation} \label{oneSas}
A_{O}=\frac{N(\Delta y>0)-N(\Delta
  y<0)}{N(\Delta y>0)+ N(\Delta y<0)}|_{P^z_{\ttbar}>P^z_c} = \frac{N(\Delta y<0)-N(\Delta
  y>0)}{N(\Delta y>0)+ N(\Delta y<0)}|_{P^z_{\ttbar}<-P^z_c}  \, , 
 \end{equation}
 where  $\Delta y=y_{t}-y_{\bar{t}}$ in the laboratory frame.

Similar to the boosted asymmetry discussed above, the fraction of the $q\bar q$
initiated $\ttbar$ sample and hence $A_{O}$ is increased by increasing
the lower cut $P^z_c$ on $P^z_{\ttbar}$. A further, more moderate  enhancement can be
achieved by applying the additional cut
  $M_{t\bar{t}} \geq M_{c}$.

In Table~\ref{tab:oscas-lhc} we collect our SM  results for $A_{O}$
 as a function of  $P^z_c$, for  $M_{c}=2 m_t$ and $0.5$ TeV,
  for the LHC at 7, 8, and 14 TeV.
 In the case of $A_{O}$ the   ratio of electroweak and QCD contributions
  remains essentially constant if $P^z_c$ and/or  $M_{c}$ is
  increased: for the results given in Table~\ref{tab:oscas-lhc} this
  ratio is between 15 and $17\%$.

The results for the one-sided  asymmetry at NLO QCD given
 in   \cite{Wang:2010du} are systematically larger than the corresponding
  QCD results given in Table~\ref{tab:oscas-lhc}. This is due to the fact that
  in \cite{Wang:2010du} the denominator of  $A_{O}$  was evaluated 
   with LO matrix elements
 and LO PDF which leads to a smaller denominator than in our case.

Finally we emphasize that the magnitudes of all LHC charge asymmetries
  discussed in this section, especially for large $\mtt$, $\beta_{min}$, or 
 $P^z_c$ cuts depend sensitively on how the denominators of the asymmetries
  are evaluated. If one computes these denominators in `Monte Carlo fashion' with
   NLO matrix elements, the magnitudes of the asymmetries decrease significantly.
   Thus the scale uncertainties given in the above tables underestimate the
  true theory uncertainties, which are rather of the order of $\sim 30\%.$

\begin{table}
\begin{centering}
\begin{tabular}{|c|c|c|c|c|c|c|c|}
\hline 
$\sqrt{s}$ & $M_{C}$ & $P_{c}^{z}$ (GeV) & 0 & 250 & 500 & 750 & 1000\tabularnewline
\hline
7 TeV & $2m_{t}$ & QCD  $A_{O}$ (\%) & 1.13~(4) & 1.50~(5) & 2.10~(6) & 2.62(8) & 3.02~(8)\tabularnewline
 &  & QCD+EW  $A_{O}$ (\%) & 1.30~(5) & 1.74~(7) & 2.43~(7) & 3.06~(9) & 3.55~(9)\tabularnewline
\cline{2-8} 
 & 500 GeV & QCD $A_{O}$ (\%) & 1.31~(4) & 1.75~(5) & 2.38~(5) & 2.95~(5) & 3.47~(4)\tabularnewline
 &  & QCD+EW  $A_{O}$ (\%) & 1.53~(5) & 2.04~(6) & 2.77~(6) & 3.45~(5) & 4.09~(3)\tabularnewline
\hline
8 TeV & $2m_{t}$ & QCD $A_{O}$ (\%) & 1.02~(4) & 1.35~(4) & 1.84~(6) & 2.39~(8) & 2.75~(9)\tabularnewline
 &  & QCD+EW  $A_{O}$ (\%) & 1.17~(5) & 1.55~(6) & 2.13~(7) & 2.78~(8) & 3.23~(9)\tabularnewline
\cline{2-8} 
 & 500 GeV & QCD  $A_{O}$ (\%) & 1.15~(3) & 1.53~(4) & 2.07~(5) & 2.61~(5) & 3.08~(6)\tabularnewline
 &  & QCD+EW  $A_{O}$ (\%) & 1.34~(4) & 1.77~(5) & 2.40~(5) & 3.04~(6) & 3.61~(5)\tabularnewline
\hline
14 TeV & $2m_{t}$ & QCD  $A_{O}$ (\%) & 0.61~(2) & 0.79~(3) & 1.07~(4) & 1.27~(6) & 1.57~(5)\tabularnewline
 &  & QCD+EW $A_{O}$ (\%) & 0.70~(5) & 0.91~(7) & 1.23~(8) & 1.48~(8) & 1.83~(9)\tabularnewline
\cline{2-8} 
 & 500 GeV & QCD $A_{O}$ (\%) & 0.74~(3) & 0.92~(3) & 1.18~(4) & 1.44~(6) & 1.74~(6)\tabularnewline
 &  & QCD+EW $A_{O}$ (\%) & 0.86~(6) & 1.07~(7) & 1.38~(9) & 1.68~(9) & 2.04~(8)\tabularnewline
\hline
\end{tabular}
\par\end{centering}
\caption{The one-sided asymmetry $A_{O}$ defined in  \eqref{oneSas}
    as a function of   $P_{c}^{z}$  without and with an additional
 cut on $\mtt$ for the LHC at 7, 8, and 14 TeV.}
\label{tab:oscas-lhc}
\end{table}


\section{Leptonic forward-backward and charge asymmetries}
\label{sec:diljet}

The asymmetries at the level of the intermediate
  $t \bar t$ states considered in the previous section cannot be measured directly, but
 are extracted from the data on  dileptonic and lepton plus jets 
 final states by an unfolding procedure. 
 On the other hand, the top-quark forward-backward and charge asymmetries lead also to 
 asymmetries for the daughter leptons from semileptonic top-quark decay. Although 
   these asymmetries are expected to be smaller than the corresponding ones for top quarks,
  because the lepton does not strictly follow the direction of its quark parent, 
  the leptonic
  asymmetries should be measurable  more precisely and should allow for a more
  direct comparison between theory and experiment. 
  
We consider here, for the Tevatron and for the LHC, 
dileptonic final states resulting from an intermediate  $\ttbar$
state:
\begin{equation}
p {\bar p}, \ p p  \rightarrow  t{\bar t}  + X 
\rightarrow \ell^+ \ell\,'^-  \, j_b \, j_{\bar b} \, + X, \label{eq:ttll}
\end{equation}
where $\ell = e,\mu$ and $j_b$ denotes a  $b$ jet.

We compute the leptonic asymmetries defined below without and with acceptance cuts. 
For the dileptonic final states we use  the following cuts ($\ell=e,\mu$,
$E^{\rm miss}_T$ denotes the missing transverse energy, and $\eta$ is the
 pseudorapidity): 
\begin{eqnarray}
{\rm Tevatron:} \quad p_T^{\ell}\geq 20 \, {\rm GeV},  \quad |\eta_{\ell}| \leq 2.0, \quad
p_T^{j}\geq 20  \,  {\rm GeV},  \quad |\eta_{j}| \leq 2.0, \quad
E^{\rm miss}_T\geq 25  \, {\rm GeV}, \label{cut:diltev} \\
{\rm LHC:} \quad p_T^{\ell }\geq 20\, {\rm GeV}, \quad
|\eta_{\ell}|\leq 2.5, \quad
p_T^{j}\geq 25\, {\rm GeV},  \quad   |\eta_{j}|\leq 2.4, \quad
E^{\rm miss}_T\geq 60 \, {\rm GeV} . \label{cut:dilhc}
\end{eqnarray}
The index $j$ refers  to a $b$, ${\bar b}$, a light (anti)quark, or
   a gluon jet, and we specify in the following how
  we apply these cuts.   

Our results for the  Tevatron asymmetry $A^\ell$ without cuts given
below apply also to the lepton plus jets events at the
Tevatron.
 \begin{equation}
p {\bar p}  \rightarrow  t{\bar t}  + X 
\rightarrow \ell^+  \, j_b \, j_{\bar b} \, j_1 \, j_2  +
X, \qquad \ell^-  \, j_{\bar b} \, j_b \, j_1 \, j_2  +
X, \label{eq:ttlj1}
\end{equation}
 where $j_{1,2}$ denote non-$b$ jets.
\subsubsection*{Tevatron}
 For the Tevatron one can  define for both 
  types of final states  \eqref{eq:ttll}, \eqref{eq:ttlj1} a
  leptonic charge asymmetry. Let
     $N_{\ell^\pm}(\eta_{\ell^\pm})$ be the number of $\ttbar$ events that
     contain a positively/negatively charged lepton $\ell^\pm$ with pseudorapidity
     $\eta_{\ell^\pm}$ in the laboratory frame. One may consider the  leptonic charge asymmetry 
 \begin{equation} \label{rapdiaslin}
   A^\ell  = \frac{N_{\ell^+}(\eta_{\ell^+}>0) \, - \, 
               N_{\ell^-}(\eta_{\ell^-}>0)} 
{ N_{\ell^+}(\eta_{\ell^+}>0) \, +  \, 
     N_{\ell^-}(\eta_{\ell^+}>0)}   \; .
\end{equation}
If CP invariance holds, then $N_{\ell^+}(\eta_{\ell^+}) = N_{\ell^-}(-\eta_{\ell^-})$ and  
 $A^\ell$ is equal to the leptonic forward-backward asymmetry,
 $A^\ell = A^{\ell^+}_{FB}  = -A^{\ell^-}_{FB}$, where
 \begin{equation} \label{rapdiaslfb}
   A^{\ell^\pm}_{FB}  = \frac{N_{\ell^\pm}(\eta_{\ell^\pm}>0) \, - \, 
      N_{\ell^\pm}(\eta_{\ell^\pm}<0)} 
{ N_{\ell^\pm}(\eta_{\ell^\pm}>0) \, +  \, 
     N_{\ell^\pm}(\eta_{\ell^\pm}<0)}   \; .
\end{equation}
In analogy to the $\tbart$ pair asymmetry $A^{\tbart}$ one may
 consider, for dileptonic final states,  the leptonic pair asymmetry
\begin{equation} 
 A^{\ell\ell} = \frac{ N_{\ell\ell}(\Delta \eta_\ell > 0) -   
 N_{\ell\ell}(\Delta \eta_\ell < 0)}{ N_{\ell\ell}(\Delta \eta_\ell > 0 ) +
  N_{\ell\ell}(\Delta \eta_\ell < 0)}\, ,
\label{fobaelelasy}
\end{equation}
where $\Delta \eta_\ell = \eta_{\ell^+}- \eta_{\ell^-}$. 
 In analogy to
 $A^{\ttbar}$ versus  $A^t_{FB}$, the pair asymmetry  $A^{\ell\ell}$ is, for kinematical
reasons,   larger than  $A^\ell$.

We  compute the asymmetries \eqref{rapdiaslfb} and
\eqref{fobaelelasy}  at NLO QCD
with respect to $\ttbar$ production and $t$ and $\bar t$ decay,
including the electroweak corrections to $\ttbar$ production as
described in Sect. \ref{suse:tevatron} and \ref{suse:lhc}. The
$\ttbar$ spin correlations are taken into account.  The
 radiative corrections were implemented into our computer code as
 described in \cite{Bernreuther:2010ny}. In computing
   the ratios \eqref{rapdiaslin} -  \eqref{fobaelelasy} we use the same procedure as in
 Sect. \ref{suse:tevatron} and \ref{suse:lhc},
 namely, we use in the denominator
   LO matrix elements  and 
            the   NLO PDF set CTEQ6.6M both in the numerator and
            denominator. 

 We  calculated the  above Tevatron observables 
  inclusively as follows. For the dileptonic 
  events, which at
 NLO in the gauge couplings  contain at most 3 partons  in the final state, we
 require that at least 2  partons satisfy the above dileptonic cuts. 
We  checked  
 that the results do not change when using 
 instead the $k_\perp$  jet algorithm \cite{Catani:1992zp}. This 
 is to say  we checked an inclusive calculation
  against $\ell^+ \nu_\ell  \ell^- {\bar \nu}_\ell \, j_b j_{\bar b}$ (LO)
   and  $\ell^+ \nu_\ell \ell^- {\bar \nu}_\ell \, j_b j_{\bar b} j$
   (NLO), where $j$ denotes a gluon  or light quark jet.

Our results for $A^\ell =
A^{\ell^+}_{FB}$ and $A^{\ell\ell}$ are collected 
 in Table~\ref{tab:llasytev}.
As expected, in the SM\footnote{The asymmetries  $A^\ell$ and   $A^{\ell\ell}$, when
   measured close to the $\ttbar$ production threshold, may contain
 information independent from the  inclusive lepton asymmetries
 \cite{Falkowski:2011zr}.}
   the leptonic charge asymmetry $A^\ell$ has the same sign
 as the  top-quark charge asymmetry given in
 Sect. \ref{suse:tevatron}, but is smaller in magnitude, while
 $A^{\ell\ell}$ is larger than  $A^\ell$ but smaller than $A^{\ttbar}$.
  As is the case for the $\ttbar$ charge asymmetry, 
 selecting events with large rapidity difference $|\Delta y_\ell|$ or large
 $\mtt$ increases  $A_\ell$ and $A^{\ell\ell}$ significantly. 

The results in the last column  of  Table~ \ref{tab:llasytev} were
obtained without applying selection cuts. 
 These numbers can be directly compared to experimental results that
 are corrected for detector effects, background contributions, and
 acceptances. As one can see from Table~\ref{tab:llasytev}, removing
 the cuts has only a minor effect on the asymmetries.

 Our results for $A^\ell$ without cuts in the last column of
 Table~\ref{tab:llasytev}
 apply also to the  lepton +
  jets final states  \eqref{eq:ttlj1} at the Tevatron.

\begin{center}
\begin{table}[h]
\begin{centering}
\begin{tabular}{|c|c|c|c|}
\hline 
 & & {with cuts} & {without cuts} \tabularnewline
\hline 
$A^{\ell}$ (\%) & QCD: & $3.0~(3)$  & $3.1~(3)$ \tabularnewline
 & QCD + EW: & $3.6~(2)$  & $3.8~(3)$ \tabularnewline
\hline
$A^{\ell}$ (\%) &   QCD: & $5.2~(5)$   & $5.8~(5)$ \tabularnewline
($\mtt \geq 450$ GeV) &  QCD + EW: &$6.4~(5)$ & $7.0~(5)$ \tabularnewline
\hline
$A^{\ell}$ (\%) & QCD: & $1.6 ~(1)$  & $1.5~(1)$ \tabularnewline
($\mtt < 450$ GeV) &  QCD + EW: & $1.9~(1)$ & $1.8~(1)$ \tabularnewline
\hline
$A^{\ell\ell}$ (\%) & QCD: & $4.0~(4)$ & $4.0~(4)$ \tabularnewline
 & QCD + EW: &  $4.8~(4)$ & $4.8~(4)$ \tabularnewline
\hline
$A^{\ell\ell}$ (\%) & QCD: & $7.0~(6)$ & $6.3~(6)$ \tabularnewline 
($|\Delta y_\ell|\geq 1$) & QCD + EW: & $8.5~(6)$ & $7.5~(6)$ \tabularnewline
\hline
$A^{\ell\ell}$ (\%) & QCD: & $1.9~(2)$ & $1.6~(1)$ \tabularnewline
($|\Delta y_\ell|<1$) & QCD + EW: & $2.3~(2)$ & $1.9~(2)$ \tabularnewline
\hline
$A^{\ell\ell}$ (\%) & QCD: & $6.7~(5)$ & $7.1~(6)$ \tabularnewline
($\mtt \geq 450$ GeV) & QCD + EW: & $8.2~(5)$ & $8.7~(6)$ \tabularnewline
\hline
$A^{\ell\ell}$ (\%) & QCD: & $2.3~(2)$ & $2.0~(2)$ \tabularnewline
($\mtt <450$ GeV) & QCD + EW:  & $2.7~(2)$ & $2.3~(2)$ \tabularnewline \hline
\end{tabular}
\par\end{centering}
\caption{The leptonic charge asymmetries  \eqref{rapdiaslfb} and
\eqref{fobaelelasy} for dileptonic final states at
   the Tevatron, computed inclusively. The
numbers in the third column were obtained by imposing the acceptance
cuts  \eqref{cut:diltev}.
  The uncertainties are due to scale variations $m_t/2 \leq \mu \leq 2
  m_t$. The results for $A^{\ell}$ without cuts apply also to lepton +
  jets final states  \eqref{eq:ttlj1}.
 }
\label{tab:llasytev}
\end{table}
\par\end{center}

\begin{center}
\begin{table}[h]
\centering{} \begin{tabular}{|c|c|c|c|}
\hline 
 & $A^{\ell}$ ~(\%) &   $A^{\ell}(\mtt< 450~{\rm GeV})$~(\%) &
 $A^{\ell}(\mtt\geq 450~{\rm GeV})$~(\%)  \tabularnewline
 \hline
 D$\emptyset$ \cite{Abazov:2011rq}  &   $15.2 \pm 4.0$ & &  \tabularnewline  \hline
 CDF  \cite{CDF2012}    &  $6.6 \pm 2.5$   &  $3.7 \pm 3.1$ & $11.6
 \pm 4.2$       \tabularnewline \hline                                   
\end{tabular}
\caption{The D$\emptyset$   \cite{Abazov:2011rq} and CDF \cite{CDF2012}  results for
  the leptonic   charge asymmetry from $\ell$ + jet events at the
  Tevatron. The  D$\emptyset$  result is  unfolded, while the CDF data
  are background-subtracted results which are not yet corrected for
  detector effects and acceptance.
}
\label{tab:D0cdf-tev}
\end{table}
\end{center}

 These asymmetries were first computed at NLO QCD (production and
 decay), including mixed QCD-weak (but not the QED)
 corrections, in \cite{Bernreuther:2010ny}. The results of 
Table~\ref{tab:llasytev} are in agreement with these results\footnote{In
   \cite{Bernreuther:2010ny} the denominators of the asymmetries were
   evaluated with LO PDF.}. In  \cite{Bernreuther:2010ny} also the
 effect of $\ttbar$ spin correlations on the leptonic asymmetries was
 investigated. Switching the spin correlations off has only a minor
 effect, which is to be expected because the
 inclusive leptonic charge asymmetries are influenced but not
 primarily caused by $t, \bar t$ spin effects.

    The asymmetry  $A^\ell$  was also
 calculated  in \cite{Bevilacqua:2010qb} for off-shell intermediate
 $t, \bar t$ at NLO QCD (production and decay, including
 non-factorizable corrections) 
 and the result of \cite{Bevilacqua:2010qb} agrees with that of
 \cite{Bernreuther:2010ny} and of Table~\ref{tab:llasytev}.
 Recently another calculation of $A^\ell$  at NLO QCD (production and
 decay),  $A^\ell=2.0^{+1.0}_{-0.3} \%$, was reported in \cite{Campbell:2012uf} for  on-shell $t, \bar
 t$. As \cite{Campbell:2012uf}
 uses the NLO QCD cross section in the denominator of  $A^\ell$, which
 is $\sim 30\%$  larger than $\sigma_{LO}$, this result is also in
 agreement with  \cite{Bernreuther:2010ny} and that of
 Table~\ref{tab:llasytev}.

So far, the D$\emptyset$ and CDF experiments have published only
results for  $A^\ell$ obtained from lepton plus jets final states,
which we have collected in Table~\ref{tab:D0cdf-tev} for the convenience 
 of the reader. While  the cited D$\emptyset$ result is the unfolded one
 \cite{Abazov:2011rq}, the CDF results  \cite{CDF2012}
  are background-subtracted but  not yet corrected for detector
  effects and acceptance. According to
  \cite{Abazov:2011rq}, unfolding  has only a
  minimal effect on the lepton asymmetry. Thus we may compare
  these experimental results with our SM predictions for $A^\ell$ in
  the no-cut case given in Table~\ref{tab:llasytev} which, as already
  mentioned above, apply also to $\ell +j$ events.
While the  CDF results agree with the SM predictions, the 
 D$\emptyset$ result $A^\ell_{\rm exp}=(15.2\pm 4)\%$ deviates by
  $\sim 2.8 \sigma.$

\subsubsection*{LHC}
 
At the LHC, leptonic charge asymmetries can  be defined for
dileptonic final states  \eqref{eq:ttll}  in analogy to the
 $\ttbar$ charge asymmetries of Sect.~\ref{suse:lhc}. 
 In analogy to  \eqref{eq:centcasy} and \eqref{edgeasy}
  we define the leptonic center and edge asymmetries
\begin{equation} \label{eq:cenasyl}
A_{C}^\ell(\eta_c) =\frac{N_{\ell\ell}(|\eta_{\ell^+}|\leq
  \eta_{c})-N_{\ell\ell}(|\eta_{\ell^-}|\leq
  \eta_{c})}{N_{\ell\ell}(|\eta_{\ell^+}|\leq
  \eta_{c})+ N_{\ell\ell}(|\eta_{\ell^-}|\leq \eta_{c})} \, , 
\end{equation}
\begin{equation} \label{edgeasyl}
A_{E}^\ell(\eta_{c})=\frac{N_{\ell\ell}(\eta_{c}\leq|\eta_{\ell^+}|)-
 N_{\ell\ell}(\eta_{c}\leq|\eta_{\ell^-}|)}{N_{\ell\ell}(\eta_{c}\leq|\eta_{\ell^+}|)
 +N_{\ell\ell}(\eta_{c}\leq|\eta_{\ell^-}|)} \, ,
\end{equation}
 where we choose in the following, for definiteness, $\eta_c=1$. \\
 The cut-independent the $\ttbar$ asymmetry \eqref{casyetalhc}
 translates to the asymmetry
\begin{equation} \label{cyetllhc}
A^{\Delta|\eta_\ell|}=\frac{N_{\ell\ell}(\Delta|\eta_\ell|>0)-
 N_{\ell\ell}(\Delta|\eta_\ell|<0)}{N_{\ell\ell}(\Delta|\eta_\ell|>0)+
 N_{\ell\ell}(\Delta|\eta_\ell|<0)} \, ,
\end{equation}
where $\Delta|\eta_\ell|=|\eta_{\ell^+}|-|\eta_{\ell^-}|$. \\
Furthermore, we define
 \begin{eqnarray} 
A_C^{\Delta|\eta_\ell|} = A^{\Delta|\eta_\ell|} \quad \text{for events
  with} \; |\Delta|\eta_\ell|| \leq\eta_c \, , \label{cccllhc} \\
A_E^{\Delta|\eta_\ell|} = A^{\Delta|\eta_\ell|} \quad \text{for events
  with} \; |\Delta|\eta_\ell|| \geq \eta_c \, , \label{eeellhc} \\
 \end{eqnarray} 
 where we choose below $\eta_c=1$, too.

\begin{center}
\begin{table}[h]
\begin{centering}
\begin{tabular}{|c|c|c|c|c|c|c|} \hline
$\sqrt{s}$ &  & $A_{C}^{\ell}$ & $A_{E}^{\ell}$ & 
 $A^{\Delta|\eta_{\ell}|}$&   $A_{C}^{\Delta|\eta_{\ell}|}$ &
$A_{E}^{\Delta|\eta_{\ell}|}$
\tabularnewline \hline 
\multicolumn{7}{|c|}{With acceptance cuts \eqref{cut:dilhc}
  (anti-$k_{T}$ and $R=0.5$)} \tabularnewline
\hline 
7 TeV & QCD \; (\%): & $-0.25~(1)$ &$0.41~(2)$ & $0.41~(2)$ &
$0.23~(1)$ & $0.95~(4)$ \tabularnewline
  &  QCD + EW \,(\%): & $-0.30~(1)$ & $0.50~(1)$ & $0.49~(1)$ &
  $0.27~(1)$ & $1.15~(2)$ \tabularnewline
\hline 
 8 TeV & QCD \; (\%): &   $-0.22~(1)$ &$0.36~(2)$ & $0.34~(2)$ &
            $0.19~(1)$ & $0.81~(3)$ \tabularnewline
 &  QCD + EW \,(\%): & $-0.27~(1)$ & $0.43~(1)$ & $0.42~(1)$ &
  $0.23~(1)$ & $0.98~(2)$ \tabularnewline
\hline 
14 TeV & QCD \; (\%): & $-0.09~(1)$ & $0.14~(1)$ & $0.09(1)$ &
$0.03~(1)$ & $0.25~(2)$\tabularnewline
       &  QCD + EW \,(\%):  & $-0.13~(1)$ &  $0.19~(1)$ & $0.14~(1)$ &
      $0.05~(2)$ &  $0.37~(2)$ \tabularnewline
\hline 
\multicolumn{7}{|c|}{Without acceptance cuts }\tabularnewline
\hline 
7 TeV & QCD \; (\%): & $-0.40~(2)$ &$0.48~(2)$ & $0.61~(3)$ &
$0.27~(2)$ & $1.25~(6)$ \tabularnewline
  &  QCD + EW \,(\%): & $-0.46~(2)$ & $0.55~(2)$ & $0.70~(3)$ &
  $0.32~(2)$ & $1.44~(6)$ \tabularnewline
\hline 
 8 TeV & QCD \; (\%): &   $-0.36~(2)$ &$0.42~(2)$ & $0.55~(3)$ &
            $0.25~(1)$ & $1.13~(6)$ \tabularnewline
 &  QCD + EW \,(\%): & $-0.42~(2)$ & $0.49~(2)$ & $0.64~(3)$ &
  $0.29~(1)$ & $1.31~(4)$ \tabularnewline
\hline 
14 TeV & QCD \; (\%): & $-0.21~(1)$ & $0.20~(1)$ & $0.36~(2)$ &
$0.15~(1)$ & $0.71~(4)$\tabularnewline
       &  QCD + EW \,(\%):  & $-0.25~(1)$ &  $0.24~(1)$ & $0.43~(2)$ &
      $0.17~(1)$ &  $0.85~(3)$ \tabularnewline
\hline 
\end{tabular}
\par\end{centering}
\caption{
  The leptonic charge asymmetries for dileptonic final states
  for the  LHC at 7, 8, and 14 TeV. The first set of numbers was
  computed using the   acceptance cuts \eqref{cut:dilhc} and  the anti-$k_T$ algorithm
  with $R=0.5$. The second set of numbers was obtained by an inclusive
  calculation without imposing cuts.  The uncertainties are due to scale variations $m_t/2 \leq \mu \leq 2
  m_t$. 
 }
\label{tab:llasy-lhc}
\end{table}
\par\end{center}

\begin{table}[h]
\begin{centering}
\begin{tabular}{|c|c|c|c|c|}
\hline 
$\sqrt{s}$ &  $A^{\Delta|\eta_{l}|}$ (\%)  & $M_{t\bar{t}}\geq 0.5$ TeV & $M_{t\bar{t}}\geq 0.7$ TeV
& $M_{t\bar{t}}\geq 1$ TeV\tabularnewline
\hline
7 TeV & QCD: & 0.94 (4) & 1.29 (3) & 1.63 (2)\tabularnewline
 & QCD+EW:    & 1.13 (2) & 1.53 (2) & 1.94 (1)\tabularnewline
\hline
8 TeV & QCD:    & 0.85 (3) & 1.16 (3) & 1.43 (9)\tabularnewline
 & QCD+EW:   & 1.03 (2) & 1.41 (1) & 1.74 (5)\tabularnewline
\hline
14 TeV & QCD: & 0.52 (2) & 0.68 (4) & 0.89 (6)\tabularnewline 
 & QCD+EW:  & 0.67 (2) & 0.88 (2) & 1.13 (3)\tabularnewline
\hline
\end{tabular}
\par\end{centering}
\caption{The leptonic charge
  asymmetry $A^{\Delta|\eta_{l}|}$, for different cuts on $M_{t\bar{t}}$,
  for dileptonic final states at the LHC without
acceptance cuts.}
\label{tab:llcutasy-lhc}
\end{table}

 In Table~\ref{tab:llasy-lhc} we collect our results for the
 leptonic charge asymmetries  \eqref{eq:cenasyl} -   \eqref{eeellhc} for dileptonic final states
  for the LHC at 7, 8, and 14 TeV. The first set of numbers was
  computed using the acceptance cuts \eqref{cut:dilhc} and  the
  anti-$k_T$ algorithm \cite{Cacciari:2008gp}
  with $R=0.5$. The second set of numbers was obtained by an inclusive
  calculation without imposing cuts.
 Our  SM results for the leptonic asymmetries follow essentially the
 same pattern that was
 found  in Sect.~\ref{suse:lhc} for the corresponding charge
 asymmetries at the level of $\ttbar$:
    $A_{C}^{\Delta|\eta_{l}|}$ is negative and smaller in magnitude
    than $A_{E}^{\ell}$ and 
 $A^{\Delta|\eta_{\ell}|}$. The edge or forward asymmetry may be
 enhanced  by the additional selection cut  \eqref{eeellhc}. 
 The $\eta_\ell$-cut independent asymmetry  $A^{\Delta|\eta_{\ell}|}$
 can be enhanced by selecting events with high pair-invariant mass
 $\mtt$, as shown in Table~\ref{tab:llcutasy-lhc}. For large $\mtt$
 the ratio of weak and QCD contributions increases somewhat.

 So far, ATLAS and CMS have not yet published results on leptonic
 asymmetries  from $\ttbar$ events. It will certainly be a challenge
 (that probably cannot be met) to detect  nonzero  effects being so
 small in magnitude than those given in
 Tables~\ref{tab:llasy-lhc},~\ref{tab:llcutasy-lhc}.
 But the point is that  these leptonic asymmetries should be excellent
 discriminators between SM and possible new physics effects, because
 it is expected that these asymmetries  will be measurable with a 
   precision of a few percent.

\section{Conclusions}
\label{sec:concl}

 We have computed  several top-quark charge
 asymmetries for $\ttbar$ production, for  the Tevatron and in
 particular for the LHC,
 to NLO QCD including mixed QCD-QED and QCD-weak interaction
 corrections. Our SM prediction for the $\ttbar$ rest-frame asymmetry
 at high mass,  $A^{\ttbar}(M_{\ttbar} >  450~{\rm GeV})$ (and those
 by other authors \cite{Hollik:2011ps,Kuhn:2011ri}) at the Tevatron deviates from the
 recent CDF measurement \cite{CDF2012} by $\sim 2.4\sigma$.
 Thus, for this observable,  the tension between experiment and SM has
 become less severe, as compared to the situation about 1 year ago \cite{Aaltonen:2011kc}. 

 For the LHC we have made SM predictions for a number of top charge
 asymmetries that were proposed in the literature, namely for the center
 and edge asymmetry, the rapidity-cut independent asymmetries (without and
 with additional cuts on $\mtt$), and for the boosted and one-sided
 asymmetry. The measurements of the inclusive  asymmetries 
 $A_{C}^{\Delta|\eta|}$,  $A_{C}^{\Delta|y|}$ by the CMS and ATLAS
 experiments agree with the SM
 results; the predictions for $\ttbar$ samples with large
 pair-invariant mass and for the other charge asymmetries still need to be
  experimentally tested. 

 Moreover, we have considered several leptonic charge asymmetries for
 dileptonic and lepton + jets  $\ttbar$ events at the Tevatron and for
  dileptonic $\ttbar$ events at the LHC. We have computed these
  asymmetries at NLO QCD (production and decay), including the mixed
  QCD-electroweak corrections to $\ttbar$ production. These leptonic
  asymmetries should be measurable rather precisely and provide
  additional information about $\ttbar$ production.
While the  CDF results on  $A^\ell$ (not yet unfolded)  agree with our predictions, the 
 unfolded D$\emptyset$ result deviates by
  $\sim 2.8 \sigma.$
  Hopefully, D$\emptyset$ and CDF will perform also  measurements of 
 the asymmetry $A^{\ell\ell}$  for dileptonic events.
 In addition, we expect that results on the LHC leptonic  charge asymmetries, for
 which we presented SM predictions in  Tables~\ref{tab:llasy-lhc}
 and~\ref{tab:llcutasy-lhc}, will become available from  ATLAS and/or
 CMS in the not too distant future.  
 The measurements of these and other distributions, 
including the search for an non-zero longitudinal polarization of the
top (anti)quarks in hadronically produced $\ttbar$ samples and
measurements of $\ttbar$ spin correlations with increased
precision\footnote{The size of the SM weak-interaction induced
 longitudinal polarization of the
top (anti)quarks was determined in \cite{Bernreuther:2006vg,Bernreuther:2008md,Bernreuther:2010ny}. Evidence and observation
of  non-zero 
 $\ttbar$ spin correlations was  recently reported by D$\emptyset$
  \cite{Abazov:2011gi} and
 by ATLAS \cite{ATLAS:2012ao}, respectively.} should eventually
clarify in detail the dynamics of hadronic $\ttbar$ production and decay.

\subsubsection*{Acknowledgements}

 We wish to thank D. Amidei, A. Harel, B. Pecjak, and C. Schwanenberger for
  discussions and correspondence. 
 The work of   W.B.  was supported by DFG, SFB TR9 and that of Z.G. Si  by NSFC and by Natural Science Foundation of
Shandong Province. 

\subsubsection*{Note added}
After submission of this manuscript several new experimental results
 appeared that are of relevance for some of our results.
 The D$\emptyset$ collaboration at the Tevatron 
  reported the measurement
of the leptonic asymmetries \eqref{rapdiaslin} --  \eqref{fobaelelasy}
 for dilepton final states and obtained the unfolded results
  \cite{:2012bf}:
 $A^\ell =(5.8 \pm 5.1({\rm stat}) \pm 1.3({\rm syst}))\%$
  and $A^{\ell\ell}=(5.3 \pm 7.9({\rm stat}) \pm 2.9({\rm syst}))\%$,
  which are in agreement with our results given in 
   Table~\ref{tab:llasytev}.
 The ATLAS collaboration 
   reported the measurement of the leptonic asymmetry
     \eqref{cyetllhc} for dilepton final states at 
 the LHC (7 TeV)~\cite{ATL12LA},
$A^{\Delta|\eta_\ell|}= (2.3 \pm 1.2({\rm stat}) \pm 0.8({\rm syst}))\%$. This measurement is in agreement with our corresponing result
   (without acceptance cuts) given in Table~\ref{tab:llasy-lhc}.
  Moreover, the recent measurements of the $\ttbar$ charge 
   asymmetry  $A_{C}^{\Delta|y|}$
   by the ATLAS \cite{ATL12LA}
  and CMS \cite{:2012xv} experiments at the LHC (7 TeV)
    are compatible with our SM prediction given 
  in Table~\ref{tab:casyy-lhc}.

\pagebreak


\begin{thebibliography}{99}



\bibitem{Aaltonen:2011kc} 
  T.~Aaltonen {\it et al.}  [CDF Collaboration],
  Phys.\ Rev.\ D {\bf 83}, 112003 (2011)
  [arXiv:1101.0034 [hep-ex]].


\bibitem{Abazov:2011rq} 
  V.~M.~Abazov {\it et al.}  [D0 Collaboration],
  Phys.\ Rev.\ D {\bf 84}, 112005 (2011)
  [arXiv:1107.4995 [hep-ex]].





\bibitem{Kuhn:1998jr} 
  J.~H.~K\"uhn and G.~Rodrigo,
  Phys.\ Rev.\ Lett.\  {\bf 81}, 49 (1998)
  [hep-ph/9802268].

\bibitem{Kuhn:1998kw}
  J.~H.~K\"uhn and G.~Rodrigo,
  Phys.\ Rev.\  D {\bf 59}, 054017  (1999) 
  [arXiv:hep-ph/9807420].

\bibitem{Bowen:2005ap}
  M.~T.~Bowen, S.~D.~Ellis and D.~Rainwater,
  Phys.\ Rev.\  D {\bf 73},  014008 (2006) 
  [arXiv:hep-ph/0509267].


\bibitem{Antunano:2007da}
  O.~Antunano, J.~H.~K\"uhn and G.~V.~Rodrigo,
  Phys.\ Rev.\  D {\bf 77},  014003  (2008) 
  [arXiv:0709.1652 [hep-ph]].



\bibitem{CDF2012} 
  T.~Aaltonen {\it et al.}  [CDF Collaboration], CDF note 10807.

\bibitem{Kamenik:2011wt} 
  J.~F.~Kamenik, J.~Shu and J.~Zupan,
  arXiv:1107.5257 [hep-ph].

\bibitem{Westhoff:2011tq}
  S.~Westhoff,
  arXiv:1108.3341 [hep-ph].



\bibitem{AguilarSaavedra:2012ma} 
  J.~A.~Aguilar-Saavedra,
  arXiv:1202.2382 [hep-ph].


\bibitem{Hollik:2011ps} 
  W.~Hollik and D.~Pagani,
  Phys.\ Rev.\ D {\bf 84}, 093003 (2011)
  [arXiv:1107.2606 [hep-ph]].



\bibitem{Bernreuther:2010ny} 
  W.~Bernreuther and Z.~-G.~Si,
  Nucl.\ Phys.\ B {\bf 837}, 90 (2010)
  [arXiv:1003.3926 [hep-ph]].



\bibitem{Chatrchyan:2011hk} 
  S.~Chatrchyan {\it et al.}  [CMS Collaboration],
  Phys.\  Lett.\  B {\bf 709}, 28 (2012)
  [arXiv:1112.5100 [hep-ex]].

\bibitem{CMSchargeas}  [CMS Collaboration], report CMS-PAS-TOP-11-030 (2012).


\bibitem{Aad:2012ug}
  G.~Aad {\it et al.}  [ATLAS Collaboration],
  arXiv:1203.4211 [hep-ex].


\bibitem{Kuhn:2011ri} 
  J.~H.~K\"uhn and G.~Rodrigo,
  JHEP {\bf 1201}, 063 (2012)
  [arXiv:1109.6830 [hep-ph]].


\bibitem{Wang:2010du} 
  Y.~-k.~Wang, B.~Xiao and S.~-h.~Zhu,
  Phys.\ Rev.\ D {\bf 82}, 094011 (2010)
  [arXiv:1008.2685 [hep-ph]].


\bibitem{Xiao:2011kp} 
  B.~Xiao, Y.~-K.~Wang, Z.~-Q.~Zhou and S.~-h.~Zhu,
  Phys.\ Rev.\ D {\bf 83}, 057503 (2011)
  [arXiv:1101.2507 [hep-ph]].

\bibitem{Hewett:2011wz} 
  J.~L.~Hewett, J.~Shelton, M.~Spannowsky, T.~M.~P.~Tait and M.~Takeuchi,
  Phys.\ Rev.\ D {\bf 84}, 054005 (2011)
  [arXiv:1103.4618 [hep-ph]].

\bibitem{Arguin:2011xm} 
  J.~-F.~Arguin, M.~Freytsis and Z.~Ligeti,
  Phys.\ Rev.\ D {\bf 84}, 071504 (2011)
  [arXiv:1107.4090 [hep-ph]].



\bibitem{AguilarSaavedra:2011cp} 
  J.~A.~Aguilar-Saavedra, A.~Juste and F.~Rubbo,
  Phys.\ Lett.\ B {\bf 707}, 92 (2012)
  [arXiv:1109.3710 [hep-ph]].



\bibitem{Alvarez:2012vq} 
  E.~Alvarez,
  arXiv:1202.6622 [hep-ph].

\bibitem{Alvarez:2012uh} 
  E.~Alvarez,
  arXiv:1205.5267 [hep-ph].


\bibitem{AguilarSaavedra:2012va}
  J.~A.~Aguilar-Saavedra and A.~Juste,
  arXiv:1205.1898 [hep-ph].




\bibitem{Bevilacqua:2010qb} 
  G.~Bevilacqua, M.~Czakon, A.~van Hameren, C.~G.~Papadopoulos and M.~Worek,
  JHEP {\bf 1102}, 083 (2011)
  [arXiv:1012.4230 [hep-ph]].





\bibitem{Dittmaier:2007wz}
  S.~Dittmaier, P.~Uwer and S.~Weinzierl,
  Phys.\ Rev.\ Lett.\  {\bf 98},  262002  (2007)
  [arXiv:hep-ph/0703120].

\bibitem{Dittmaier:2008uj}
  S.~Dittmaier, P.~Uwer and S.~Weinzierl,
  Eur.\ Phys.\ J.\  C {\bf 59},   625 (2009)
  [arXiv:0810.0452 [hep-ph]].


\bibitem{Melnikov:2010iu} 
  K.~Melnikov and M.~Schulze,
  Nucl.\ Phys.\ B {\bf 840}, 129 (2010)
  [arXiv:1004.3284 [hep-ph]].

\bibitem{Melnikov:2011qx} 
  K.~Melnikov, A.~Scharf and M.~Schulze,
  Phys.\ Rev.\ D {\bf 85}, 054002 (2012)
  [arXiv:1111.4991 [hep-ph]].

\bibitem{Alioli:2011as} 
  S.~Alioli, S.~-O.~Moch and P.~Uwer,
  JHEP {\bf 1201}, 137 (2012)
  [arXiv:1110.5251 [hep-ph]].



\bibitem{Almeida:2008ug}
  L.~G.~Almeida, G.~Sterman and W.~Vogelsang,
  Phys.\ Rev.\  D {\bf 78}, 014008  (2008) 
  [arXiv:0805.1885 [hep-ph]].

\bibitem{Ahrens:2010zv} 
  V.~Ahrens, A.~Ferroglia, M.~Neubert, B.~D.~Pecjak and L.~L.~Yang,
  JHEP {\bf 1009}, 097 (2010)
  [arXiv:1003.5827 [hep-ph]].

\bibitem{Kidonakis:2011zn} 
  N.~Kidonakis,
  Phys.\ Rev.\ D {\bf 84}, 011504 (2011)
  [arXiv:1105.5167 [hep-ph]].



\bibitem{Ahrens:2011uf} 
  V.~Ahrens, A.~Ferroglia, M.~Neubert, B.~D.~Pecjak and L.~L.~Yang,
  Phys.\ Rev.\ D {\bf 84}, 074004 (2011)
  [arXiv:1106.6051 [hep-ph]].




\bibitem{Campbell:2012uf}
  J.~M.~Campbell and R.~K.~Ellis,
  arXiv:1204.1513 [hep-ph].



\bibitem{Frixione:2008ym}
  S.~Frixione and B.~R.~Webber,
  arXiv:0812.0770 [hep-ph].



 
\bibitem{Frixione:2007nw}
  S.~Frixione, P.~Nason and G.~Ridolfi,
  JHEP {\bf 0709}, 126  (2007) 
  [arXiv:0707.3088 [hep-ph]].



\bibitem{MCFM}
J.~Campbell and  R.~K.~Ellis, {\tt http:/mcfm.fnal.gov}

\bibitem{Campbell:2010ff} 
  J.~M.~Campbell and R.~K.~Ellis,
  Nucl.\ Phys.\ Proc.\ Suppl.\  {\bf 205-206}, 10 (2010)
  [arXiv:1007.3492 [hep-ph]].








\bibitem{Bernreuther:2005is}
  W.~Bernreuther, M.~Fuecker and Z.~G.~Si,
  Phys.\ Lett.\  B {\bf 633},  54 (2006)
  [arXiv:hep-ph/0508091].

\bibitem{Bernreuther:2006vg}
  W.~Bernreuther, M.~Fuecker and Z.~G.~Si,
  Phys.\ Rev.\  D {\bf 74},   113005 (2006)
  [arXiv:hep-ph/0610334].

\bibitem{Bernreuther:2008md}
  W.~Bernreuther, M.~F\"ucker and Z.~G.~Si,
  Phys.\ Rev.\  D {\bf 78},  017503  (2008)
  [arXiv:0804.1237 [hep-ph]].

\bibitem{Kuhn:2005it}
  J.~H.~K\"uhn, A.~Scharf and P.~Uwer,
  Eur.\ Phys.\ J.\  C {\bf 45}, 139  (2006)
  [arXiv:hep-ph/0508092].

\bibitem{Kuhn:2006vh}
  J.~H.~K\"uhn, A.~Scharf and P.~Uwer,
  Eur.\ Phys.\ J.\  C {\bf 51},  37 (2007)
  [arXiv:hep-ph/0610335].


\bibitem{Beenakker:1993yr}
  W.~Beenakker, A.~Denner, W.~Hollik, R.~Mertig, T.~Sack and D.~Wackeroth,
  Nucl.\ Phys.\  B {\bf 411},  343 (1994).

\bibitem{Brodsky:2012ik} 
  S.~J.~Brodsky and X.~-G.~Wu,
  arXiv:1205.1232 [hep-ph].



\bibitem{Nadolsky:2008zw}
  P.~M.~Nadolsky, H.~-L.~Lai, Q.~-H.~Cao, J.~Huston, J.~Pumplin, D.~Stump, W.~-K.~Tung and C.~-P.~Yuan,
  Phys.\ Rev.\ D {\bf 78},  013004 (2008) 
  [arXiv:0802.0007 [hep-ph]].



\bibitem{Martin:2009iq}
  A.~D.~Martin, W.~J.~Stirling, R.~S.~Thorne and G.~Watt,
  Eur.\ Phys.\ J.\  C {\bf 63},  189  (2009)
  [arXiv:0901.0002 [hep-ph]].




\bibitem{Catani:1992zp}
  S.~Catani, Y.~L.~Dokshitzer and B.~R.~Webber,
  Phys.\ Lett.\  B {\bf 285}, 291  (1992).




\bibitem{Cacciari:2008gp}
  M.~Cacciari, G.~P.~Salam and G.~Soyez,
  JHEP {\bf 0804},  063  (2008)
  [arXiv:0802.1189 [hep-ph]].




\bibitem{Falkowski:2011zr}
  A.~Falkowski, G.~Perez and M.~Schmaltz,
  arXiv:1110.3796 [hep-ph].


\bibitem{Abazov:2011gi}
  V.~M.~Abazov {\it et al.}  [D0 Collaboration],
  Phys.\ Rev.\ Lett.\  {\bf 108},  032004 (2012) 
  [arXiv:1110.4194 [hep-ex]].

\bibitem{ATLAS:2012ao}
  G.~Aad {\it et al.}  [ATLAS Collaboration],
  arXiv:1203.4081 [hep-ex].

\bibitem{Drobnak:2012cz} 
  J.~Drobnak, J.~F.~Kamenik and J.~Zupan,
arXiv:1205.4721 [hep-ph]. 

\bibitem{:2012bf} 
  V.~M.~Abazov {\it et al.}  [D0 Collaboration],
 arXiv:1207.0364 [hep-ex]. 



\bibitem{ATL12LA}
G. Aad et al. [ATLAS Collaboration], report ATLAS-CONF-2012-057.

\bibitem{:2012xv} 
  S.~Chatrchyan {\it et al.}  [CMS Collaboration],
  arXiv:1207.0065 [hep-ex]. 

\end{thebibliography}
\end{document}